\documentclass[12pt, draftclsnofoot, onecolumn]{IEEEtran}
\IEEEoverridecommandlockouts
\usepackage{cite}
\usepackage{bbm}
\usepackage{amsmath,amssymb,amsfonts,amsthm}
\usepackage{graphicx}
\usepackage{textcomp}
\usepackage{xcolor}
\usepackage{authblk}
\usepackage{stfloats}
\usepackage{soul}
\theoremstyle{definition}

\usepackage{cleveref}
\usepackage{subcaption}
\usepackage{mathtools}
\usepackage[normalem]{ulem}
\usepackage{algorithm}
\usepackage{algpseudocode}

\def\BibTeX{{\rm B\kern-.05em{\sc i\kern-.025em b}\kern-.08em
    T\kern-.1667em\lower.7ex\hbox{E}\kern-.125emX}}

\makeatletter
\newcommand\footnoteref[1]{\protected@xdef\@thefnmark{\ref{#1}}\@footnotemark}
\makeatother

\begin{document}
\newcounter{cor_cnt}
\setcounter{cor_cnt}{0}
\newtheorem{cor}{Corollary}
\bstctlcite{IEEEexample:BSTcontrol}
\title{A Reduced Complexity Ungerboeck Receiver for Quantized Wideband Massive SC-MIMO}
\author{A. Bulut \"{U}\c{c}\"{u}nc\"{u},~\IEEEmembership{Student Member,~IEEE,}
{G\"{o}khan M. G\"{u}vensen,~\IEEEmembership{Member,~IEEE,}}
{and~A. \"{O}zg\"{u}r Y{\i}lmaz,~\IEEEmembership{Member,~IEEE}}
\thanks{The work of A. B. \"{U}\c{c}\"{u}nc\"{u} was supported by
the ASELSAN Graduate Scholarship for Turkish Academicians. \newline\indent
A. B. \"{U}\c{c}\"{u}nc\"{u}, G. M. G\"{u}vensen and A. \"{O}. Y{\i}lmaz are with the Department
of Electrical and Electronics Engineering, Middle East Technical University, 06800, Ankara,
Turkey, (e-mail: \{ucuncu, guvensen, aoyilmaz\}@metu.edu.tr)}
}


\maketitle

%

\begin{abstract}
Employing low resolution analog-to-digital converters in massive multiple-input multiple-output (MIMO) has many advantages in terms of total power consumption, cost and feasibility of such systems. However, such advantages come together with significant challenges in channel estimation and data detection due to the severe quantization noise present. In this study, we propose a novel iterative receiver for quantized uplink single carrier MIMO (SC-MIMO) utilizing an efficient message passing algorithm based on the Bussgang decomposition and Ungerboeck factorization, which avoids the use of a complex whitening filter. A reduced state sequence estimator with bidirectional decision feedback is also derived, achieving remarkable complexity reduction compared to the existing receivers for quantized SC-MIMO in the literature, without any requirement on the sparsity of the transmission channel. Moreover, the linear minimum mean-square-error (LMMSE) channel estimator for SC-MIMO under frequency-selective channel, which do not require any cyclic-prefix overhead, is also derived. We observe that the proposed receiver has significant performance gains with respect to the existing receivers in the literature under imperfect channel state information.
\end{abstract}

\section{Introduction}
\label{sec:intro}
Massive multiple-input multiple-output (MIMO) systems have been enjoying considerable attention to be deployed in modern communication systems due to the many advantages they provide regarding spectral and energy efficiency \cite{larsson_MM_survey,swindlehurst_overview,ICC_own,Marzetta1088544,TCOMM_ACI_own}. 
Despite their advantages, owing to the large number of antennas, power consumption and cost of the components per antenna becomes a hindrance. Regarding this bottleneck, utilization of low-resolution analog-to-digital converters (ADCs) at each antenna of the MIMO array shines out as a feasible solution, due to their lower power consumption and cost \cite{VTC_own,jacobsson_recent,jacobsson}. Therefore, their utilization in massive MIMO systems have been examined intensively \cite{TWC_own,Globecom_ACI_own,VTC_seq_lin_rec,chan_est_flat_1,chan_est_flat_2,chan_est_flat_3,chan_est_flat_4,
flat_fading_det_14,jacobsson,larsson_MM_1bitADC}. 
However, the aforementioned advantages of low resolution ADCs are not for free; there is an increased difficulty for two major tasks in such systems, namely the channel estimation and data detection, due to quantization distortion. There are many studies in the literature that handles these two important problems. In this work, we propose a novel data detector and a channel estimator for quantized wideband single-carrier (SC) massive uplink MIMO and achieve significant advantages compared to the existing work in the literature.
\subsection{Related Work}
Regarding channel estimation algorithms in quantized MIMO, there are numerous works \cite{chan_est_flat_1,chan_est_flat_2,chan_est_flat_3,chan_est_flat_4,
flat_fading_det_14,jacobsson,larsson_MM_1bitADC,chan_est_flat_8,chan_est_flat_9,
chan_est_flat_10,chan_est_flat_11,chan_est_flat_12,
chan_est_flat_13,chan_est_flat_14,chan_est_flat_15,chan_est_flat_16,chan_est_flat_17,
chan_est_flat_18,chan_est_flat_19,chan_est_flat_20,chan_est_flat_21,
chan_est_freq_sel_1,chan_est_freq_sel_2,
Mollen_wideband,Li_perf_analysis,chan_est_freq_sel_3,chan_est_freq_sel_6,chan_est_freq_sel_7,chan_est_freq_sel_8,chan_est_freq_sel_9}. Among those studies \cite{chan_est_flat_1,chan_est_flat_2,chan_est_flat_3,chan_est_flat_4,
flat_fading_det_14,jacobsson,larsson_MM_1bitADC,chan_est_flat_8,chan_est_flat_9,
chan_est_flat_10,chan_est_flat_11,chan_est_flat_12,
chan_est_flat_13,chan_est_flat_14,chan_est_flat_15,chan_est_flat_16,chan_est_flat_17,
chan_est_flat_18,chan_est_flat_19,chan_est_flat_20,chan_est_flat_21} propose channel estimation algorithms for flat fading channels. As quantization is a non-linear operation, the extension of flat fading channel estimation techniques to frequency-selective channels is not straightforward. 

There are also many works that propose various channel estimation algorithms for frequency-selective channels in quantized MIMO systems \cite{chan_est_freq_sel_1,chan_est_freq_sel_2,
Mollen_wideband,Li_perf_analysis,chan_est_freq_sel_3,chan_est_freq_sel_6,chan_est_freq_sel_7,
chan_est_freq_sel_8,chan_est_freq_sel_9}. Among them, \cite{chan_est_freq_sel_1} proposes a channel estimation method to estimate sparse frequency-selective channels for orthogonal frequency division multiplexing (OFDM) modulation. Another study \cite{chan_est_freq_sel_2} combines the expectation-maximization (EM) algorithm with a channel estimation method relying on the sparsity in the channel. However, the complexity of the algorithm in \cite{chan_est_freq_sel_2} can be high as it involves the maximization of $M$ non-linear, non-convex functions over $KL$ complex variables, $M$ being the number of antennas, $K$ and $L$ being the number of users and channel taps. Due to the non-convex nature of the channel estimation problem, the global optimum solution is also not guaranteed.

More recently, \cite{Mollen_wideband} proposed a linear low-complexity channel estimation technique for frequency-selective channels in for MIMO-OFDM, without any sparsity assumption on the channel, where the quantization noise is assumed to be an independent identically distributed noise. This assumption is only accurate as the number of taps or users becomes large. In contrast, \cite{Li_perf_analysis} takes into account the correlation between the quantization noise terms by deriving the linear minimum mean-square-error (LMMSE) channel estimate. Another study \cite{chan_est_freq_sel_3} proposes a low complexity channel estimation algorithm based on approximate message-passing, showing some performance improvement compared to LMMSE channel estimation. However, the aforementioned channel estimation techniques \cite{Mollen_wideband,Li_perf_analysis,chan_est_freq_sel_3} require OFDM and a cyclic-prefix (CP), which may decrease the spectral efficiency significantly, especially if $L$ is large. The same OFDM or CP limitation also exists for the channel estimation techniques in \cite{chan_est_freq_sel_6,chan_est_freq_sel_3,chan_est_freq_sel_7,chan_est_freq_sel_8,chan_est_freq_sel_9,SVM_swindlehurst}.
In short, for all of the aforementioned channel estimation methods for frequency-selective fading, at least one of the following limitations holds: the requirement of OFDM or a CP \cite{chan_est_freq_sel_1,Mollen_wideband,Li_perf_analysis,chan_est_freq_sel_3,chan_est_freq_sel_6,chan_est_freq_sel_7,chan_est_freq_sel_8,chan_est_freq_sel_9}, the requirement of a sparse channel \cite{chan_est_freq_sel_1,chan_est_freq_sel_6,chan_est_freq_sel_7,chan_est_freq_sel_8,chan_est_freq_sel_9}, or being computationally very complex \cite{chan_est_freq_sel_2}. 

For the channel estimation algorithm to be proposed in our study, we concentrate on a linear and low complexity method in quantized massive MIMO systems under frequency selective fading, which does not require any CP and can work with single-carrier (SC) modulation. The reason for the selection of SC modulation is owing to the fact that SC is superior to OFDM for systems having nonlinearities, such as quantized MIMO \cite{phase_only,chan_est_freq_sel_9}, owing to its lower peak-to-average power ratio (PAPR) \cite{papr_sc_ofdm} and robustness to carrier-frequency-offset (CFO) errors \cite{single_car_vs_ofdm}.

Regarding data detection in quantized massive MIMO, there are also a vast number of studies in the literature \cite{flat_fading_det_1,flat_fading_det_2,
flat_fading_det_3,chan_est_flat_21,flat_fading_det_5,flat_fading_det_6,
flat_fading_det_7,flat_fading_det_8,flat_fading_det_9,flat_fading_det_10,
chan_est_flat_8,flat_fading_det_12,flat_fading_det_14,
flat_fading_det_15,flat_fading_det_16,Studer_ML_optim,ref_138,
SVM_swindlehurst,ref_131,phase_only,gampı_bitiren_makale,gamp_iceren_arxiv_makale,ref_133,dinnis_benchmark_sc_fde,
heat_sphere_det,robust_freq_selective_sparsity_simo_youseb,CP_free,
MLSE_MA_MIMO}. Among them, \cite{flat_fading_det_1,flat_fading_det_2,
flat_fading_det_3,chan_est_flat_21,flat_fading_det_5,flat_fading_det_6,
flat_fading_det_7,flat_fading_det_8,flat_fading_det_9,flat_fading_det_10,
chan_est_flat_8,flat_fading_det_12,flat_fading_det_14,
flat_fading_det_15,flat_fading_det_16} propose detectors for frequency-flat channels. However, for wideband transmission, frequency-flat channel assumption is not practical, even with mmWave channels \cite{rappaport}.

For frequency-selective channels, \cite{Studer_ML_optim,ref_138,SVM_swindlehurst,ref_131,
phase_only,gampı_bitiren_makale,gamp_iceren_arxiv_makale,ref_133,dinnis_benchmark_sc_fde,
heat_sphere_det,robust_freq_selective_sparsity_simo_youseb,CP_free,
MLSE_MA_MIMO} advocate various data detectors for quantized massive MIMO systems. Among those work, \cite{Studer_ML_optim} proposes a maximum \textit{a posteriori} (MAP) detection for OFDM. The complexity of the proposed MAP algorithm is very high, thus suboptimal but lower complexity versions are also proposed in \cite{Studer_ML_optim}. However, their performance is shown to be even worse than a relatively much lower complexity per subcarrier LMMSE data equalization method \cite{robust_freq_selective_sparsity_simo_youseb}. Another study \cite{ref_138} employs a generalized approximate message passing (GAMP) based detector to obtain the Bayes-optimal data estimates for quantized MIMO. However, if OFDM is not employed, the proposed detector requires the inversion of $N_c\times N_c$ matrices, $N_c$ being the number of subcarriers, which means a complexity growth with $N_c^3$, while the complexity of the detector proposed in our work grows linearly with $N_c$. Even with OFDM, its complexity grows with $M^2N_c$ \cite{ref_138}, which indicates a computationally exhaustive detector. A much recent study \cite{SVM_swindlehurst} also proposes a support vector machine based detector for one-bit massive MIMO systems, which is again limited to OFDM, requiring a CP. Moreover, the complexity of the data detection in \cite{SVM_swindlehurst} grows with $N_c^2K^2$ for frequency selective channels. 

Owing to the aforementioned advantages of SC systems over OFDM, there are many studies proposing SC frequency domain equalization (SC-FDE) detectors in quantized MIMO. To start with, \cite{ref_131} proposes a GAMP based detector. However, as the proposed algorithm cannot be applied to arbitrary constellations, \cite{ref_131} is later extended to work with arbitrary constellations in \cite{phase_only} and \cite{gampı_bitiren_makale}. However, the number of nonlinear operations per iteration of the proposed receivers in \cite{phase_only} and \cite{gampı_bitiren_makale} is $5MNO+PKN$, where $O$ is a number between $80$ and $100$, $P$ is the modulation order and $N$ is the data packet length, which can be compared to $N_c$ of the multi-carrier modulation schemes. As the number of antennas $M$ in massive MIMO is large, $5MNO$ becomes a very large number. Moreover, the number of iterations for GAMP based methods to converge is typically about 10 iterations \cite{ref_138}, whereas the proposed detector in this work will be observed to converge in about $2$ iterations in most cases. This means a prohibitive complexity for the detectors proposed in \cite{phase_only} and \cite{gampı_bitiren_makale} for massive MIMO. Moreover, \cite{gamp_iceren_arxiv_makale} also advocates a GAMP based receiver, but its complexity grows with $N^2$, which can also be very high. Another SC-FDE and GAMP based detector is proposed in \cite{ref_133}. Nevertheless, the detector in \cite{ref_133} is limited to spatial modulation, which is not a commonly used technique. More recently, \cite{dinnis_benchmark_sc_fde} proposed various iterative detectors with feasible complexity for quantized massive MIMO.

Despite being superior compared to OFDM for quantized MIMO, there is still a CP overhead in SC-FDE. Therefore, some recent work has proposed detectors that can work without a CP for quantized single carrier MIMO (SC-MIMO) systems under frequency-selective fading. A sphere decoding based detector is proposed in \cite{heat_sphere_det} for one-bit massive MIMO. However, the complexity of the detector in \cite{heat_sphere_det} increases with $MN^3KP$ for frequency-selective fading case, which is computationally infeasible. Moreover, \cite{robust_freq_selective_sparsity_simo_youseb} proposes a reinforcement learning based data detector. Nonetheless, $L$ being the number of channel taps, the computational complexity of the detector in \cite{robust_freq_selective_sparsity_simo_youseb} grows with $P^L$ \cite{simo_önceki_makale}, which is only feasible for extremely sparse channels. In contrast, the computational complexity grows linearly with $L$ for the proposed data detector in our work. Another detector is also proposed in \cite{CP_free} as a frequency domain equalizer. However, its complexity grows with $NK^3$, while the proposed detector in this work has a complexity growth with $NK^2$. More recently, a maximum-likelihood sequence estimator for one-bit wideband massive MIMO is proposed in \cite{MLSE_MA_MIMO}. The computational complexity of the detector in \cite{MLSE_MA_MIMO} grows with $P^L$, resulting in excessive complexity for large $L$. In the same study, the necessity of a decision feedback equalization based reduced state detector is also mentioned as a future work, which is one of the important properties of the proposed detector in our study.
\subsection{Contributions}
In this work, we propose a novel iterative detector for quantized SC uplink MIMO, which will be referred to as QA-UMPA-BDF, standing for quantization-aware Ungerboeck type message passing algorithm with bidirectional decision feedback. To incorporate the quantization effects on the observation model we utilize Bussgang decomposition. An efficient message passing structure is proposed based on Ungerboeck factorization, through which we also avoid the need to use a complex noise whitening filter \cite{gmguvensen_2014_reduced_state}. To further reduce the computational complexity, a reduced state sequence estimator with bidirectional decision feedback is also derived. To sum up, the main contribution items associated with our paper are as follows:
\begin{itemize}
\item{The proposed detector is one of the few detectors in the literature that are able to work without any CP in quantized massive SC-MIMO, which can be important for a feasible spectral efficiency. In the literature, the detectors working without CP has a complexity growth with $P^L$, whereas the complexity of the proposed detector grows linearly with $L$.}
\item{The proposed detector is derived not only for one-bit ADCs but also for multi-bit ADCs.}
\item{The proposed detector is the first reduced state Ungerboeck type detector with bidirectional decision feedback structure working in wideband MIMO even for the unquantized case.}
\item{The proposed detector provides significant error-rate performance advantages over the higher complexity representative detector \cite{dinnis_benchmark_sc_fde} from the literature, which even has lower spectral efficiency due to the lack of CP in our detection scheme.}
\item{LMMSE channel estimates for CP-free SC quantized MIMO is derived based on Bussgang decomposition, which does not exist in the literature up to the knowledge of the authors.}
\end{itemize}
As the benchmark detector to compare with the proposed detector, we prefer the detectors designed for SC modulation to OFDM based detectors, as SC modulation is superior to OFDM when a non-linearity is present, due to its lower PAPR \cite{papr_sc_ofdm} and robustness to CFO errors \cite{single_car_vs_ofdm}. Among the aforementioned SC detectors, the ones that have comparable complexity to our detector, when the number of channel taps are not very low (if the channel is not extremely sparse), are \cite{dinnis_benchmark_sc_fde} and \cite{CP_free}. We prefer \cite{dinnis_benchmark_sc_fde} over \cite{CP_free} as the benchmark algorithm even if \cite{dinnis_benchmark_sc_fde} has a CP overhead, since \cite{dinnis_benchmark_sc_fde} is a much more recent study and the performance of \cite{CP_free} may be inferior to \cite{dinnis_benchmark_sc_fde} due to the inter-carrier interference caused by the lack of CP in \cite{CP_free}.

\textit{Notation:} $c$ is a scalar, $\mathbf{c}$ is a column vector, $\mathbf{C}$ is a matrix and $\mathbf{C}^H$, $\mathbf{C}^T$, and $\mathbf{C}^*$ represent the Hermitian, transpose, and conjugate of matrix $\mathbf{C}$, respectively. $[\mathbf{C}]_{(m,n)}$ stands for the element of matrix $\mathbf{C}$ at its $m^{th}$ row and $n^{th}$ column. $\mathbb{E}[.]$ takes the expectation of its operand.  $\operatorname{Re}(.)$ and $\operatorname{Im}(.)$ take the real and imaginary parts of their operands and $j=\sqrt{-1}$. $||.||$ corresponds to the Euclidean norm. $\mathbf{0}_K$ and $\mathbf{I}_K$ are zero and identity matrices with size $K\times K$; and $\mathrm{diag}(\mathbf{C})$ is the diagonal matrix, whose diagonal entries are equal to the diagonal entries of $\mathbf{C}$. Moreover, $\mathrm{log}_2(.)$ is the base-2 logarithm, $\otimes$ is the Kronecker product and $\text{Tr}[.]$ is the trace operator. Furthermore, $\mathcal{Q}(.)$ is the Q-function, which is defined as $\mathcal{Q}(x)=\frac{1}{\sqrt{2\pi}}\int_{x}^{\infty}e^{-z^2/2}dz$. The notation $\mathrm{blkToeplitz}(\mathbf{C},\mathbf{R})$ indicates a block Toeplitz matrix of dimension $M\times K$, whose first row block is matrix $\mathbf{R}$ of size $M_r\times K$ and the first column block is matrix $\mathbf{C}$ of size $M\times K_c$.
\section{System Model}
\begin{figure}
\centering
\includegraphics[width=0.80\columnwidth]{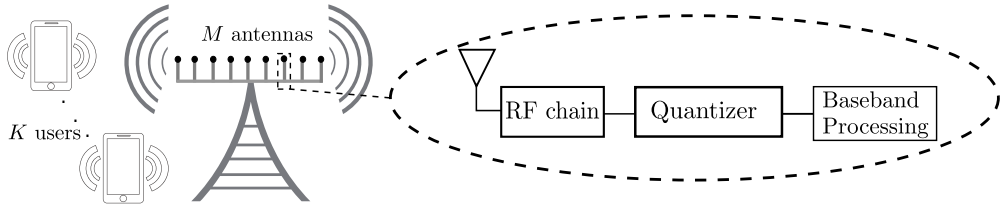}
\caption{Uplink massive MIMO system with low resolution quantizers.}
\label{fig:syst_model_ma_MIMO}
\end{figure}

In this study, a frequency-selective single-cell uplink massive MIMO system with $K$ single-antenna users and $M$ receive antennas with low-resolution ADCs as in Fig~\ref{fig:syst_model_ma_MIMO} is examined.
In this case, the unquantized received signal at the $m^{th}$ antenna $y_m[n]$ can be expressed as
\begin{equation}
\label{eqn:rec_sig_discr_time}
y_m[n]=\sum_{k=1}^{K}\sum_{\ell=0}^{L-1}\sqrt{\rho_{k}[\ell]}h_{m,k}[\ell]x_k[n-\ell] + w_m[n],
\end{equation}
where $L$ is the number of channel taps, $h_{m,k}[\ell]$ is the $\ell^{th}$ tap in the impulse response of the channel between the $k^{th}$ user and the $m^{th}$ receive antenna, $w_m[n]$ is the thermal noise sample received from the $m^{th}$ antenna at the $n^{th}$ sample time. Moreover, $\rho_{k}[\ell]$ is the power-delay profile of the channel between the receive antennas and the $k^{th}$ user, satisfying $\sum_{\ell=1}^{L}\rho_{k}[\ell]=1, \forall k$. The channel taps $h_{m,k}[\ell]$ are assumed to be zero-mean unit variance circularly symmetric complex Gaussian (CSCG) random variables, corresponding to a Rayleigh fading scenario, and uncorrelated, that is, $\mathbb{E}[h_{m_1,k_1}[\ell_1]h_{m_2,k_2}[\ell_2]^*]=\delta[\ell_1-\ell_2]\delta[k_1-k_2]\delta[m_1-m_2]$. Such assumptions for the channel coefficients are commonly adopted in many studies \cite{Mollen_wideband,jacobsson_aggragate,jacobsson_lin_prec}, among others.  
Thermal noise samples are assumed to be independent identically distributed (i.i.d) zero-mean CSCG random variables with variance $N_o$. Moreover, $x_k[n]$ is the transmitted symbol by user $k$ at the $n^{th}$ time index, with average symbol energy $E_s=\mathbb{E}[|x_k[n]|^2], \forall k,n$. (\ref{eqn:rec_sig_discr_time}) can be rewritten compactly as
\begin{equation}
\label{eqn:rec_sig_discr_time_mtx_vec}
\mathbf{y}[n]=\sum_{\ell=0}^{L-1}\mathbf{H}[\ell]\mathbf{J}[\ell]\mathbf{x}[n-\ell] + \mathbf{w}[n],
\end{equation}
where $\mathbf{J}[\ell]$ is a $K\times K$ diagonal matrix, whose $k^{th}$ diagonal is $\sqrt{\rho_k[\ell]}$, and $\mathbf{H}[\ell]$ is the MIMO channel matrix, whose element at its $m^{th}$ row and the $k^{th}$ column is equal to $h_{m,k}[\ell]$. Moreover, $\mathbf{x}[n]$ and $\mathbf{w}[n]$ are the transmitted symbol and noise vectors, whose $k^{th}$ and $m^{th}$ elements are equal to $x_k[n]$ and $w_m[n]$, respectively. The quantized received signal can also be expressed as
\begin{equation}
\label{eqn:quantized signal}
\mathbf{r}[n]=\mathrm{Q}(\mathbf{y}[n]),
\end{equation}
where $\mathrm{Q}(.)$ is the function mapping the input of the quantizer to its output. For the case of 1-bit quantizer,  $\mathrm{Q}(.)=\mathrm{sign}(\operatorname{Re}(.))+j\mathrm{sign}(\operatorname{Im}(.))$, $\mathrm{sign}(.)$ being the sign function.

\section{LMMSE Channel Estimation for CP-free Quantized SC-MIMO}
In this section, the expression for the LMMSE channel estimate for quantized SC-MIMO systems will be derived. In the channel estimation phase, we assume that each user transmits pilot signals simultaneously. In this case, the received signal in (\ref{eqn:rec_sig_discr_time_mtx_vec}) can be reexpressed as
\begin{align}
\label{eqn:unquantized rec_sig_pilot}
\underline{\mathbf{y}}^{(p)}&=\left(\mathbf{X}\otimes \mathbf{I}_M\right)\underline{\mathbf{h}}+\underline{\mathbf{w}},\\
\underline{\mathbf{y}}^{(p)}\triangleq\left[\mathbf{y}[0]^T \ \mathbf{y}[1]^T \cdots \mathbf{y}[\tau-1]^T\right]^T, \  \underline{\mathbf{w}}\triangleq&\left[\mathbf{w}[0]^T \ \mathbf{w}[1]^T \cdots \mathbf{w}[\tau-1]^T\right]^T, \mathbf{X}\triangleq\left[\mathbf{X}_1 \ \mathbf{X}_2 \cdots \mathbf{X}_K \right], \nonumber \\\underline{\mathbf{h}}\triangleq\left[\left({\mathbf{h}}^{(1)}\right)^T \ \left({\mathbf{h}}^{(2)}\right)^T \cdots \left({\mathbf{h}}^{(K)}\right)^T \right]^T, &\
{\mathbf{h}}^{(k)}\triangleq\left[\left({\mathbf{h}}^{(k)}[0]\right)^T \ \left({\mathbf{h}}^{(k)}[1]\right)^T \cdots \left({\mathbf{h}}^{(k)}[L-1]\right)^T \right]^T, \nonumber
\end{align}
in which $\tau$ is training length, ${\mathbf{h}}^{(k)}[\ell]$ is the $k^{th}$ column of $\mathbf{H}[\ell]$, $\left[\mathbf{X}_k\right]_{(m+1,n+1)}\triangleq\sqrt{\rho_k[n]}x_k[m-n]$, where $x_k[m]$, $m=0,1,\ldots,\tau-1$, is the transmitted pilot of user $k$, and $n=0,1,\ldots,L-1$.
\subsection{Bussgang Decomposition based LMMSE Channel Estimator}
To obtain a linear and simple channel estimator, we utilize the Bussgang decomposition \cite{bussgang_decomp}, which enables finding a statistically equivalent linear operator for any nonlinear function \cite{ozlem_bussgang}. According to the Bussgang decomposition, $\underline{\mathbf{r}}^{(p)}=\mathrm{Q}(\underline{\mathbf{y}}^{(p)})$ can be written as
\begin{equation}
\label{eqn:quant_rec_sig_pilot}
\underline{\mathbf{r}}^{(p)}=\underline{\mathbf{A}}^{(p)}\underline{\mathbf{y}}^{(p)}+\underline{\mathbf{q}}^{(p)}.
\end{equation}
We denote the cross-covariance matrix between $\underline{\mathbf{y}}^{(p)}$ and $\underline{\mathbf{r}}^{(p)}$ by $\mathbf{C}_{\underline{\mathbf{y}}^{(p)}\underline{\mathbf{r}}^{(p)}}$ and the autocovariance matrix of ${\underline{\mathbf{y}}^{(p)}}$ by $\mathbf{C}_{\underline{\mathbf{y}}^{(p)}}$. When $\mathbf{A}^{(p)}$ is selected as $\mathbf{A}^{(p)}=\mathbf{C}_{\underline{\mathbf{y}}^{(p)}\underline{\mathbf{r}}^{(p)}}^H \mathbf{C}_{\underline{\mathbf{y}}^{(p)}}^{-1}$, the distortion term $\underline{\mathbf{q}}^{(p)}$ is minimized, equivalently, $\underline{\mathbf{q}}^{(p)}$ is made uncorrelated with $\underline{\mathbf{y}}^{(p)}$. For a one-bit quantizer, assuming zero-mean Gaussian inputs\footnote{\label{ftnt:Quant_inp_gauss}This assumption is approximately true even when the transmitted symbols are not Gaussian but from a finite cardinality set. The input of the ADC at the $m^{th}$ antenna can be written as a sum of $KL$ i.i.d. random variables with finite variance. Owing to the central limit theorem (CLT), the value of this sum (ADC input) converge to Gaussian as $KL$ grows large.}, the following holds \cite{TCOMM_ACI_own}:
\begin{align}
\label{eqn:a_mtx_exp}
\underline{\mathbf{A}}^{(p)}=\sqrt{4/\pi}\mathrm{diag}(\mathbf{C}_{\underline{\mathbf{y}}^{(p)}})^{-0.5}&=\sqrt{4/\pi}\mathrm{diag}\left(\left(\mathbf{X}\mathbf{X}^H\otimes \mathbf{I}_M\right)+N_o\mathbf{I}_{M\tau}\right)^{-0.5}\\
\label{eqn:a_mtx_exp2}
&=\mathbf{B}\otimes \mathbf{I}_M,
\end{align}
where $\mathbf{B}=\sqrt{4/\pi}\mathrm{diag}(\mathbf{X}\mathbf{X}^H+N_o\mathbf{I}_\tau)^{-0.5}$.
Moreover, the auto-covariance matrix of the distortion term at the quantizer output $\underline{\mathbf{q}}^{(p)}$, namely $\mathbf{\mathbf{C}_{\underline{\mathbf{q}}^{(p)}}}$, can be expressed using (\ref{eqn:quant_rec_sig_pilot}) as
\begin{equation}
\label{eqn:quant_dist_long_cov}
\mathbf{C}_{\underline{\mathbf{q}}^{(p)}}=\mathbf{C}_{\underline{\mathbf{r}}^{(p)}}-\underline{\mathbf{A}}^{(p)}\mathbf{C}_{\underline{\mathbf{y}}^{(p)}}\left(\underline{\mathbf{A}}^{(p)}\right)^H,
\end{equation}
where $\mathbf{C}_{\underline{\mathbf{r}}^{(p)}}=\mathbb{E}[\underline{\mathbf{r}}^{(p)}\left(\underline{\mathbf{r}}^{(p)}\right)^H]$ can be found using \textit{arcsine law} \cite{van1966spectrum} for one-bit quantizer as
\begin{equation}
\label{eqn:arcsine_rule}
\mathbf{C}_{\underline{\mathbf{r}}^{(p)}}=\dfrac{4}{\pi}\left(\mathrm{asin}\left(\mathbf{D_{\underline{\mathbf{y}}^{(p)}}}^{-\frac{1}{2}}\operatorname{Re}(\mathbf{C_{\underline{\mathbf{y}}^{(p)}}})\mathbf{D_{\underline{\mathbf{y}}^{(p)}}}^{-\frac{1}{2}}\right)+j \ \mathrm{asin}\left(\mathbf{D_{\underline{\mathbf{y}}^{(p)}}}^{-\frac{1}{2}}\operatorname{Im}(\mathbf{C_{\underline{\mathbf{y}}^{(p)}}})\mathbf{D_{\underline{\mathbf{y}}^{(p)}}}^{-\frac{1}{2}}\right)\right),
\end{equation}
where $\mathbf{D}_{\underline{\mathbf{y}}^{(p)}}=\mathrm{diag}\left(\mathbf{C_{\underline{\mathbf{y}}^{(p)}}}\right)$. Plugging $\mathbf{C_{\underline{\mathbf{y}}^{(p)}}}=\left(\mathbf{X}\mathbf{X}^H\otimes \mathbf{I}_M\right)+N_o\mathbf{I}_{M\tau}=\left(\mathbf{X}\mathbf{X}^H+N_o\mathbf{I}_\tau\right)\otimes \mathbf{I}_M$ in (\ref{eqn:arcsine_rule}), one can find that
\begin{equation}
\label{eqn:arcsine_rule2}
\mathbf{C}_{\underline{\mathbf{r}}^{(p)}}=\mathbf{C}_{\mathbf{\eta}}\otimes \mathbf{I}_M,
\end{equation}
\begin{equation}
\label{eqn:arcsine_rule3}
\mathbf{C}_{\mathbf{\eta}}=\dfrac{4}{\pi}\left(\mathrm{asin}\left(\mathbf{K_{\underline{\mathbf{y}}^{(p)}}}^{-\frac{1}{2}}\operatorname{Re}(\mathbf{G_{\underline{\mathbf{y}}^{(p)}}})\mathbf{K_{\underline{\mathbf{y}}^{(p)}}}^{-\frac{1}{2}}\right)+j \ \mathrm{asin}\left(\mathbf{K_{\underline{\mathbf{y}}^{(p)}}}^{-\frac{1}{2}}\operatorname{Im}(\mathbf{G_{\underline{\mathbf{y}}^{(p)}}})\mathbf{K_{\underline{\mathbf{y}}^{(p)}}}^{-\frac{1}{2}}\right)\right),
\end{equation}
$\mathbf{G}_{\underline{\mathbf{y}}^{(p)}}=\left(\mathbf{X}\mathbf{X}^H+N_o\mathbf{I}_\tau\right)$, $\mathbf{K}_{\underline{\mathbf{y}}^{(p)}}=\mathrm{diag}\left(\mathbf{G}_{\underline{\mathbf{y}}^{(p)}}\right)$. Then, it follows from (\ref{eqn:a_mtx_exp2}), (\ref{eqn:quant_dist_long_cov}) and (\ref{eqn:arcsine_rule2}) that 
\begin{equation}
\label{eqn:distortion_cov_mtx_simple}
\mathbf{C}_{\underline{\mathbf{q}}^{(p)}}=\mathbf{E}\otimes \mathbf{I}_M,
\end{equation}
where $\mathbf{E}=\mathbf{C}_\eta-\mathbf{B}\left(\mathbf{X}\mathbf{X}^H+N_o\mathbf{I}_\tau\right)\mathbf{B}^H$. Now that some simple expressions are found for $\underline{\mathbf{A}}^{(p)}$ and $\mathbf{C}_{\underline{\mathbf{q}}^{(p)}}$ in (\ref{eqn:a_mtx_exp2}) and (\ref{eqn:distortion_cov_mtx_simple}), the quantized signal $\underline{\mathbf{r}}^{(p)}$ can be found using (\ref{eqn:unquantized rec_sig_pilot}),  (\ref{eqn:quant_rec_sig_pilot}) and (\ref{eqn:a_mtx_exp2}) as
\begin{align}
\label{eqn:quant_obs}
\underline{\mathbf{r}}^{(p)}=\left(\mathbf{B}\mathbf{X}\otimes \mathbf{I}_M\right)\underline{\mathbf{h}}+\left(\mathbf{B}\otimes \mathbf{I}_M\right)\underline{\mathbf{w}}+\underline{\mathbf{q}}^{(p)}.
\end{align} 
Let $\underline{\mathbf{\Gamma}}\triangleq\left(\mathbf{B}\otimes \mathbf{I}_M\right)\underline{\mathbf{w}}+\underline{\mathbf{q}}^{(p)}$, which corresponds to the total effective noise at the quantizer output. Its covariance matrix $\mathbf{C}_{\underline{\mathbf{\Gamma}}}$ can be found using (\ref{eqn:distortion_cov_mtx_simple}) and (\ref{eqn:quant_obs}) as
\begin{equation}
\label{eqn:eff_noise_cov}
\mathbf{C}_{\underline{\mathbf{\Gamma}}}=\mathbf{F}\otimes\mathbf{I}_M,
\end{equation}
where $\mathbf{F}=N_o\mathbf{B}\mathbf{B}^H+\mathbf{E}$. Now that the effective noise covariance matrix is found, we can apply a whitening filter, $\mathbf{C}_{\underline{\mathbf{\Gamma}}}^{-1/2}=\mathbf{F}^{-1/2}\otimes \mathbf{I}_M$, to obtain
\begin{equation}
\label{eqn:whitened_obs}
\underline{\mathbf{z}}^{(p)}\triangleq\mathbf{C}_{\underline{\mathbf{\Gamma}}}^{-1/2}\underline{\mathbf{r}}^{(p)}=\left(\mathbf{P}\mathbf{X}\otimes \mathbf{I}_M\right)\underline{\mathbf{h}}+\underline{\mathbf{n}},
\end{equation}
where $\mathbf{P}=\mathbf{F}^{-1/2}\mathbf{B}$ and $\underline{\mathbf{n}}=\left(\mathbf{P}\otimes \mathbf{I}_M\right)\underline{\mathbf{w}}+\mathbf{C}_{\underline{\mathbf{\Gamma}}}^{-1/2}\underline{\mathbf{q}}^{(p)}$, whose covariance matrix $\mathbf{C}_{\underline{\mathbf{n}}}=\mathbf{I}_{M\tau}$. To derive the LMMSE estimator from the whitened observations, we also need to find whether $\underline{\mathbf{h}}$ and $\underline{\mathbf{n}}$ are uncorrelated. It has been shown in \cite[Appendix A]{Li_perf_analysis} that for any quantized LMMSE channel estimation based on Bussgang decomposition, the quantization distortion term, which is $\underline{\mathbf{q}}^{(p)}$ in our work, is uncorrelated with the channel estimates, implying that $\underline{\mathbf{n}}$ is also uncorrelated with $\underline{\mathbf{h}}$ as $\underline{\mathbf{w}}$ is also uncorrelated with $\underline{\mathbf{h}}$. In this case, $\mathbf{C}_{\underline{\mathbf{z}}^{(p)}}\triangleq\mathbb{E}[{\underline{\mathbf{z}}^{(p)}}{\underline{\mathbf{z}}^{(p)}}^H]$ can be found as
\begin{equation}
\label{eqn:whit_obs_autocov}
\mathbf{C}_{\underline{\mathbf{z}}^{(p)}}=\left(\mathbf{X'}\mathbf{X'}^H\otimes \mathbf{I}_M\right)+\mathbf{I}_{M\tau},
\end{equation}
where $\mathbf{X'}=\mathbf{P}\mathbf{X}$.
$\mathbf{C}_{\underline{\mathbf{z}}^{(p)}\underline{\mathbf{h}}}\triangleq\mathbb{E}\left[\underline{\mathbf{z}}^{(p)}\underline{\mathbf{h}}^H\right]$ can also be obtained as
\begin{align}
\label{eqn:cross_corr}
\mathbf{C}_{\underline{\mathbf{z}}^{(p)}\underline{\mathbf{h}}}=\left(\mathbf{X}'\otimes \mathbf{I}_M\right).
\end{align}
Consequently, the LMMSE channel estimate for CP-free wideband one-bit massive SC-MIMO, namely $\hat{\underline{\mathbf{h}}}^{\text{LMMSE}}$, can be found using (\ref{eqn:whit_obs_autocov}) and (\ref{eqn:cross_corr}) as
\begin{align}
\label{eqn:LMMSE_chan_est}
\hat{\underline{\mathbf{h}}}^{\text{LMMSE}}=\mathbf{C}_{\underline{\mathbf{z}}^{(p)}\underline{\mathbf{h}}}^{H}\mathbf{C}_{\underline{\mathbf{z}}^{(p)}}^{-1}\underline{\mathbf{z}}^{(p)}=
\left(\mathbf{X}'\otimes \mathbf{I}_M\right)^H\left(\left(\mathbf{X'}\mathbf{X'}^H\otimes \mathbf{I}_M\right)+\mathbf{I}_{M\tau}\right)^{-1}\underline{\mathbf{z}}^{(p)}.
\end{align}
Note that in (\ref{eqn:LMMSE_chan_est}), the inverse of a $M\tau\times M\tau$ matrix should be taken. To obtain a lower complexity LMMSE estimator, we define $\mathbf{X''}\triangleq\mathbf{X'}\otimes \mathbf{I}_M$. Then, (\ref{eqn:LMMSE_chan_est}) can be rewritten as
\begin{align}
\hat{\underline{\mathbf{h}}}^{\text{LMMSE}}=\mathbf{X''}^H\left(\mathbf{X''}\mathbf{X''}^H+\mathbf{I}_{M\tau}\right)^{-1}\underline{\mathbf{z}}^{(p)}.
\label{eqn:LMMSE_chan_est_2}
\end{align}
Employing Woodbury matrix identity \cite{brown2012practical}, (\ref{eqn:LMMSE_chan_est_2}) can be reexpressed as
\begin{align}
\hat{\underline{\mathbf{h}}}^{\text{LMMSE}}&=\left(\left(\mathbf{X''}\right)^H\mathbf{X''}+\mathbf{I}_{MKL}\right)^{-1}\left(\mathbf{X''}\right)^H\underline{\mathbf{z}}^{(p)}\nonumber\\&=\left(\left(\left(\mathbf{X'}\right)^H\mathbf{X'}\otimes \mathbf{I}_M\right)+\mathbf{I}_{MKL}\right)^{-1}\left(\mathbf{X''}\right)^H\underline{\mathbf{z}}^{(p)}\nonumber\\&=\left(\left(\mathbf{X}^H\mathbf{P}^H\mathbf{PX}+\mathbf{I}_{KL}\right)^{-1}\otimes \mathbf{I}_M\right)\left(\mathbf{PX}\otimes \mathbf{I}_M\right)^H\underline{\mathbf{z}}^{(p)}.
\label{eqn:LMMSE_chan_est_3}
\end{align}
In (\ref{eqn:LMMSE_chan_est_3}), the inverse of a $KL\times KL$ matrix is taken, much less complex than taking the inverse of an $M\tau \times M\tau$ matrix in (\ref{eqn:LMMSE_chan_est}), as $\tau\geq KL$ in general (for orthogonal pilot assignment to users $\tau\geq KL$).  This complexity, along with the complexity to obtain $\mathbf{F}^{-1/2}$, which is $\mathcal{O}\left(\tau^3\right)$, is much less than the one derived in \cite{Li_perf_analysis} for OFDM, which requires CP decreasing spectral efficiency and taking the inverse of an $M\tau\times M\tau$ matrix, which is large as $M$ is large in massive MIMO and $\tau\geq KL$ in general. Mean-square-error (MSE) matrix for $\hat{\underline{\mathbf{h}}}^{\text{LMMSE}}$ can also be calculated as
\begin{align}
\mathbf{C}_{\underline{\mathbf{\hat{h}}}}^{\text{LMMSE}}&=\mathbb{E}\left[\left(\hat{\underline{\mathbf{h}}}^{\text{LMMSE}}-{\underline{\mathbf{h}}}^{\text{LMMSE}}\right)\left(\hat{\underline{\mathbf{h}}}^{\text{LMMSE}}-{\underline{\mathbf{h}}}^{\text{LMMSE}}\right)^H\right]\nonumber\\
&=\mathbf{I}_{MKL}-\left(\left(\mathbf{X}^H\mathbf{P}^H\mathbf{PX}+\mathbf{I}_{KL}\right)^{-1}\otimes \mathbf{I}_M\right)\left(\mathbf{X}^H\mathbf{P}^H\mathbf{PX}\otimes \mathbf{I}_M\right).
\label{eqn:mse}
\end{align}

\subsection{Low Complexity Approximations for the LMMSE Estimator}
\label{sec:low_comp_lmmse}
In this section, we will show that an even lower complexity approximation for LMMSE channel estimate exists under some conditions. One of those conditions is $\mathbf{X}\mathbf{X}^H$ being diagonally dominant. This happens when the pilots assigned to different users are nearly orthogonal and the autocorrelation function of the transmitted pilot sequence of all users is close to an impulse function, that is $\sum_{n=0}^{L-1}\sqrt{\rho_k[n]}\sqrt{\rho_{k'}[n]}x_k[m-n]x_{k'}^*[m'-n]\approx U\delta[m-m']\delta[k-k']$, where $U$ is a multiplicative constant. This approximation is accurate when users transmit randomly generated complex symbols as pilot sequences and $KL$ is large. Another condition for LMMSE channel estimate to have a lower complexity approximation is signal-to-noise ratio (SNR) being low. For both cases (for low SNR or when $\mathbf{X}\mathbf{X}^H$ is diagonally dominant) $\mathbf{C_{\underline{\mathbf{y}}^{(p)}}}=\left(\mathbf{X}\mathbf{X}^H\otimes \mathbf{I}_M\right)+N_o\mathbf{I}_{M\tau}$ is diagonally dominant, in which case $\mathbf{C}_{\underline{\mathbf{q}}^{(p)}}$ can be approximated using (\ref{eqn:quant_dist_long_cov})-(\ref{eqn:arcsine_rule3}) as
\begin{equation}
\label{eqn:white_noise_approx}
\mathbf{C}_{\underline{\mathbf{q}}^{(p)}}\approx \left(2-4/\pi\right)\mathbf{I}_{M\tau}.
\end{equation}
In that case, the overall effective noise becomes uncorrelated. Then, the complexity of the calculation of the whitening filter $\mathbf{C}_{\underline{\mathbf{\Gamma}}}^{-1/2}=\mathbf{F}^{-1/2}\otimes \mathbf{I}_M$ is reduced significantly as $\mathbf{F}$ is now a diagonal matrix for uncorrelated effective noise. This makes $\mathbf{P}=\mathbf{F}^{-1/2}\mathbf{B}$ a diagonal matrix as $\mathbf{B}$ is a diagonal matrix by definition, reducing the complexity in calculating (\ref{eqn:LMMSE_chan_est_3}). 

A further reduction in the complexity of the LMMSE channel estimator is possible by assuming that $\mathbf{B}=\sqrt{4/\pi}\mathrm{diag}(\mathbf{X}\mathbf{X}^H+N_o\mathbf{I}_\tau)^{-0.5}$ is a constant diagonal matrix. This is an accurate assumption when SNR is low. Even if SNR is not low, it is valid to assume that the diagonal elements of $\mathbf{X}\mathbf{X}^H$, corresponding to the average received power for each received sample at each antenna (averaged over channel realizations), which are equal to $\sum_{\ell=0}^{L-1}\sum_{k=0}^{K-1}\rho_k[l]|x_k[m-\ell]|^2$ for all receive antennas at the $m^{th}$ received sample, does not change over the pilot symbol transmission phase for most cases. If the magnitude of the transmitted complex pilot symbols are always the same, which is the case for DFT pilot sequences or any sequence generated randomly from a phase-shift keying  (PSK) type modulation, this assumption is exactly correct. Otherwise, the approximation error caused by this assumption goes to zero as $KL$ becomes large. With this assumption, $\mathbf{B}$ can be approximated as
\begin{equation}
\label{eqn:approx_B}
\mathbf{B}=\sqrt{4/\pi}\mathrm{diag}(\mathbf{X}\mathbf{X}^H+N_o\mathbf{I}_\tau)^{-0.5}\approx g\mathbf{I}_\tau,
\end{equation}
where $g=\sqrt{4/{(\pi(KE_s+N_o))}}$. Along with (\ref{eqn:white_noise_approx}), this implies that
\begin{equation}
\label{eqn:approx_P}
\mathbf{P}=\mathbf{F}^{(-1/2)}\mathbf{B}\approx(2-4/\pi+g^2N_o)^{(-1/2)}g\mathbf{I}_{\tau}.
\end{equation}
Then, the LMMSE estimator in (\ref{eqn:LMMSE_chan_est_3}) and MSE expression in (\ref{eqn:mse}) can be approximated as
\begin{equation}
\label{eqn:LMMSE_white_noise}
\hat{\underline{\mathbf{h}}}^{\text{LMMSE}}\approx\left(\left(c^2\mathbf{X}^H\mathbf{X}+\mathbf{I}_{KL}\right)^{-1}\otimes \mathbf{I}_M\right)\left(c\mathbf{X}\otimes \mathbf{I}_M\right)^H\underline{\mathbf{z}}^{(p)},
\end{equation}
\begin{equation}
\label{eqn:mse_white_noise}
\mathbf{C}_{\underline{\mathbf{\hat{h}}}}^{\text{LMMSE}}\approx\mathbf{I}_{MKL}-\left(\left(c^2\mathbf{X}^H\mathbf{X}+\mathbf{I}_{KL}\right)^{-1}\otimes \mathbf{I}_M\right)\left(c^2\mathbf{X}^H\mathbf{X}\otimes \mathbf{I}_M\right),
\end{equation}
where $c=g/\sqrt{2-4/\pi+g^2N_o}$. By approximating $\mathbf{X}\mathbf{X}^{H}$ as a constant diagonal matrix (with diagonal entries being $KE_s+N_o$), which is an accurate assumption under the conditions mentioned above, expressions of even lower complexity to calculate can be written from (\ref{eqn:LMMSE_chan_est}) as
\begin{equation}
\label{eqn:LMMSE_white_noise_2}
\hat{\underline{\mathbf{h}}}^{\text{LMMSE}}\approx\dfrac{c\left(\mathbf{X}\otimes \mathbf{I}_M\right)^H\underline{\mathbf{z}}^{(p)}}{c^2E_sK+1}, \mathbf{C}_{\underline{\mathbf{\hat{h}}}}^{\text{LMMSE}}\approx\left(1-\dfrac{c^2\left(\mathbf{X}^H\mathbf{X}\otimes\mathbf{I}_M\right)}{c^2E_sK+1}\right),
\end{equation}
which are very simple expressions not involving any matrix inversions.
\subsection{Extension to multi-bit quantizers}
The LMMSE channel estimator and the resulting MSE expression for multi-bit quantizers are the same as (\ref{eqn:LMMSE_chan_est_3}) and (\ref{eqn:mse}), except that the matrices $\mathbf{B}$ and $\mathbf{E}$ employed in the derivation of (\ref{eqn:LMMSE_chan_est_3}) and (\ref{eqn:mse}) are modified. For the example case of multi-bit midrise uniform quantizers with Gaussian inputs\footnote{See footnote \ref{ftnt:Quant_inp_gauss}.}, $\mathbf{B}$ in (\ref{eqn:a_mtx_exp2}) can be found as \cite{jacobsson_aggragate},
\begin{align}
\mathbf{B}=\frac{\Delta}{\sqrt{\pi}}\mathrm{diag}\left(\mathbf{X}\mathbf{X}^H+N_o\mathbf{I}_\tau\right)^{-0.5}\sum_{i=1}^{2^{q}-1}\mathrm{exp}\left(-{\Delta}^2\left(i-2^{q-1}\right)^2\mathrm{diag}\left(\mathbf{X}\mathbf{X}^H+N_o\mathbf{I}_\tau\right)^{-0.5}\right),\label{eqn:mtx_A_for_multi_bit}
\end{align}
where $\Delta$ is the quantizer step size and $q$ is the number of quantizer bits. Moreover, $\mathbf{E}$ in (\ref{eqn:distortion_cov_mtx_simple}) can be approximated for multi-bit quantizer case as \cite{jacobsson_aggragate}
\begin{align}
\label{eqn:multi_bit_noise_covar}
\mathbf{E}\approx &\dfrac{\Delta^2}{2}(2^q-1)^2\mathbf{I}_{\tau}-\mathbf{B}\mathrm{diag}\left(\mathbf{X}\mathbf{X}^H+N_o\mathbf{I}_\tau\right)\mathbf{B}^H\nonumber\\&-4\Delta^2\sum_{i=1}^{2^q-1}\left(i-2^{q-1}\right)\times\left(1-\mathcal{Q}\left(\sqrt{2}(i-2^{q-1})\mathrm{diag}\left(\mathbf{X}\mathbf{X}^H+N_o\mathbf{I}_\tau\right)^{-1/2}\right)\right).
\end{align}
Moreover, for the lower complexity LMMSE channel estimator calculations in Section \ref{sec:low_comp_lmmse}, replacing $\mathbf{X}\mathbf{X}^{H}$ by a constant diagonal matrix with diagonal entries being $(KE_s)$ and defining $P_r\triangleq KE_s+N_o$, the following approximations can be used for the multi-bit quantizer case 
\begin{align}
g=\sqrt{\Delta^2/\left(\pi P_r \right)}\times \sum_{i=1}^{2^{q}-1}&\mathrm{exp}\left(-{\Delta}^2\left(i-2^{q-1}\right)^2/\sqrt{\left(P_r \right)}\right), c\approx g/\sqrt{(d+g^2N_o)},
\label{eqn:mtx_A_for_multi_bit_approx}
\end{align}
\begin{align}
\label{eqn:multi_bit_noise_covar_approx}
d=\dfrac{\Delta^2}{2}(2^q-1)^2-g^2(P_r)-4\Delta^2\sum_{i=1}^{2^q-1}\left(i-2^{q-1}\right)\times\left(1-\mathcal{Q}\left(\sqrt{2}(i-2^{q-1})\left(P_r \right)^{-1/2}\right)\right).
\end{align}
\section{Data Transmission}
For the data transmission phase, the quantized received signal can be rewritten using (\ref{eqn:rec_sig_discr_time_mtx_vec}) as
\begin{equation}
\label{eqn:unquantized_rec_sig_data}
\underline{\mathbf{r}}^{(d)}=\mathrm{Q}\left(\underline{\mathbf{y}}^{(d)}\right)=\mathrm{Q}\left({\mathbf{\underline{H}}}\hspace{2pt}{\mathbf{\underline{{x}}}}+\mathbf{\underline{w}}\right),
\end{equation}
\begin{align}
\underline{\mathbf{y}}^{(d)}\triangleq\left[\mathbf{y}[0]^T \ \mathbf{y}[1]^T \cdots \mathbf{y}[N+L-2]^T\right]^T, \  \underline{\mathbf{w}}&\triangleq\left[\mathbf{w}[0]^T \ \mathbf{w}[1]^T \cdots \mathbf{w}[N+L-2]^T\right]^T,\nonumber
\end{align}
\begin{align}
\underline{\mathbf{H}}\triangleq\mathrm{blkToeplitz}(\underline{\mathbf{H}}^c,\underline{\mathbf{H}}^r), \ \underline{\mathbf{x}}\triangleq\left[\mathbf{x}[0]^T \ \mathbf{x}[1]^T \cdots \mathbf{x}[N-1]^T\right]^T,
\end{align}
in which $\underline{\mathbf{H}}^c\triangleq\left[\mathbf{\tilde{H}}[0]^T \ \mathbf{\tilde{H}}[1]^T \ \cdots \ \mathbf{\tilde{H}}[L-1]^T \ \mathbf{0} \cdots \mathbf{0}\right]^T$ with size $(N+L-2)M\times K$ and $\underline{\mathbf{H}}^r\triangleq\left[\mathbf{\tilde{H}}[0] \ \mathbf{0} \ \cdots \ \mathbf{0}\right]$ with size $M\times NK$, where
$\mathbf{\tilde{H}}[\ell]=\mathbf{{H}}[\ell]\mathbf{J}[\ell]$. Note that despite the same or similar notations are used for the data/pilot and noise vectors for the channel estimation and data transmission signal models for simplicity, they are completely independent of each other. Using Bussgang decomposition \cite{bussgang_decomp}, (\ref{eqn:unquantized_rec_sig_data}) can be reexpressed as
\begin{equation}
\label{eqn:data_rec_sig_bussg}
\underline{\mathbf{r}}^{(d)}=\underline{\mathbf{A}}^{(d)}{\mathbf{\underline{H}}}\hspace{2pt}{\mathbf{\underline{{x}}}}+\underline{\mathbf{A}}^{(d)}\mathbf{\underline{w}}+\mathbf{\underline{q}}^{(d)},
\end{equation}
where $\underline{\mathbf{A}}^{(d)}$ can be found by replacing $\mathbf{C}_{\underline{\mathbf{y}}^{(p)}}$ in (\ref{eqn:a_mtx_exp}) by $\mathbf{C}_{\underline{\mathbf{y}}^{(d)}}\triangleq E_s{\mathbf{\underline{H}}}\hspace{1pt}{\mathbf{\underline{H}}}^H+N_o\mathbf{I}_{M\left(N+L-2\right)}$ for one-bit quantizer or by replacing $\mathbf{X}\mathbf{X}^H+N_o\mathbf{I}_{\tau}$ in (\ref{eqn:mtx_A_for_multi_bit}) by $\mathbf{C}_{\underline{\mathbf{y}}^{(d)}}$
for multi-bit uniform midrise quantizer. Moreover, the quantizer distortion term covariance matrix, namely $\mathbf{C}_{\mathbf{\underline{q}}^{(d)}}$, can be found for one-bit quantizer case by replacing $\mathbf{C}_{\underline{\mathbf{y}}^{(p)}}$, $\mathbf{C}_{\underline{\mathbf{r}}^{(p)}}$ and $\underline{\mathbf{A}}^{(p)}$ in (\ref{eqn:quant_dist_long_cov}) and (\ref{eqn:arcsine_rule}) by $\mathbf{C}_{\underline{\mathbf{y}}^{(d)}}$, $\mathbf{C}_{\underline{\mathbf{r}}^{(d)}}\triangleq\mathbb{E}[{\underline{\mathbf{r}}^{(d)}}{\underline{\mathbf{r}}^{(d)}}^H]$ and $\underline{\mathbf{A}}^{(d)}$, respectively. For multi-bit quantizer case, $\mathbf{C}_{\mathbf{\underline{q}}^{(d)}}$ can also be obtained by replacing $\mathbf{X}\mathbf{X}^H+N_o\mathbf{I}_{\tau}$ and $\mathbf{I}_{\tau}$ in (\ref{eqn:multi_bit_noise_covar}) by $\mathbf{C}_{\underline{\mathbf{y}}^{(d)}}$ and $\mathbf{I}_{M(N+L-2)}$, respectively. Moreover, as the channel coefficients are taken to be i.i.d. unit variance random variables, all diagonal elements of $\mathbf{H}\mathbf{H}^H$, which correspond to received average signal power, converge to $KE_s$ as $KL$ goes large, with a similar discussion made previously for the pilot transmission phase. Moreover, when $KL$ is large or for low SNR, again it is straightforward to show that $\mathbf{H}\mathbf{H}^H+N_o\mathbf{I}_{M(N+L-2)}$ is a diagonally dominant matrix with diagonal entries converging to $KE_s+N_o$. In this case (when $\mathbf{H}\mathbf{H}^H+N_o\mathbf{I}_{M(N+L-2)}\approx (KE_s+N_o)\mathbf{I}_{M(N+L-2)}$), by employing the modified versions of (\ref{eqn:a_mtx_exp}), (\ref{eqn:quant_dist_long_cov}), (\ref{eqn:arcsine_rule}),  (\ref{eqn:mtx_A_for_multi_bit}), (\ref{eqn:multi_bit_noise_covar}) for data transmission phase (with the aforementioned modifications such as replacing $\mathbf{C}_{\underline{\mathbf{y}}^{(p)}}$ by $\mathbf{C}_{\underline{\mathbf{y}}^{(d)}}$), $\underline{\mathbf{A}}^{(d)}$ and $\mathbf{C}_{\underline{\mathbf{q}}^{(d)}}$ can be approximated as
\begin{equation}
\label{eqn:approx_data_detection}
\underline{\mathbf{A}}^{(d)}\approx a\mathbf{I}_{M(N+L-2)},\hspace{15pt}
\mathbf{C}_{\underline{\mathbf{q}}^{(d)}}\approx e\mathbf{I}_{M(N+L-2)},
\end{equation}
where $a=\sqrt{4/(\pi(KE_s+N_o))}$, $e={2-4/\pi}$ for one-bit quantizer case. For multi-bit quantizer case $a=g$, where $g$ can be found using (\ref{eqn:mtx_A_for_multi_bit_approx}) and $e=d$, where $d$ can be found using (\ref{eqn:multi_bit_noise_covar_approx}). The aforementioned assumptions, leading to an uncorrelated quantizer noise assumption, are observed to be accurate even when the number of users are as low as $K=4$ and $L=1$ for i.i.d. channel coefficients \cite[Fig.~4]{Li_perf_analysis}, even for high SNR. In fact, $K$ and $L$ values will be much larger in general, implying very low approximation errors.

\section{Quantization Aware Ungerboeck Type Message Passing Algorithm with Bidirectional Decision Feedback}
Based on (\ref{eqn:data_rec_sig_bussg}) and (\ref{eqn:approx_data_detection}), a minimum distance performance metric can be constructed as
\begin{equation}
\label{eqn:min_dist_metric}
\Lambda\left(\underline{\underline{\mathbf{r}}},\underline{\underline{\mathbf{H}}},\underline{\mathbf{x}}\right)=\gamma_1\mathrm{exp}\left(-||\underline{\underline{\mathbf{r}}}-\underline{\underline{\mathbf{H}}}\hspace{1pt}\underline{\mathbf{x}}||^2\right),
\end{equation}
where $\underline{\underline{\mathbf{r}}}={\underline{\mathbf{r}}}/\sqrt{e+a^2N_o}$, $\underline{\underline{\mathbf{H}}}=a{\underline{\mathbf{H}}}/\sqrt{e+a^2N_o}$ and $\gamma_1$ is a multiplicative constant. The metric corresponds to the ML metric when the effective noise term $(a\mathbf{\underline{w}})+\mathbf{\underline{q}^{(d)}}$ has a Gaussian distribution. It has been pointed out in \cite{Mezghani_cap_LB,erkip_low_power,bai2015energy} that the Gaussian assumption for the effective noise $(a\mathbf{\underline{w}})+\mathbf{\underline{q}^{(d)}}$ yields accurate results, especially for low SNR, even for 1-bit quantizer. With this finding, as $a$ increases to get closer to 1 and the power of quantizer noise $\mathbf{\underline{q}^{(d)}}$ decreases with increasing number of bits, it can be stated that the effective noise $(a\mathbf{\underline{w}})+\mathbf{\underline{q}^{(d)}}$ can be approximated as Gaussian also for higher quantization resolutions, as the $(a\mathbf{\underline{w}})$ term in the effective noise dominates. Therefore, there are many studies that approximates the quantization noise as Gaussian \cite{Mezghani_cap_LB,erkip_low_power,bai2015energy,
gauss_approx_1,mazo1968quantizing,bennett_gauss,deviceinc}.
We continue by rewriting the minimum distance metric in (\ref{eqn:min_dist_metric}) as
\begin{align}
\Lambda\left(\underline{\underline{\mathbf{r}}},\underline{\underline{\mathbf{H}}},\underline{\mathbf{x}}\right)&=\gamma_2\mathrm{exp}\left(2\Re\left(\underline{\underline{\mathbf{r}}}^H\underline{\underline{\mathbf{H}}}\hspace{2pt}\underline{\mathbf{x}}\right)-\underline{\mathbf{x}}^H\underline{\underline{\mathbf{H}}}^H\underline{\underline{\mathbf{H}}}\hspace{2pt}\underline{\mathbf{x}}\right).
\label{eqn:min_dist_metric_contd}
\end{align}
To obtain the optimal estimates based on the metric in (\ref{eqn:min_dist_metric_contd}), there are various approaches. One is to filter $\underline{\underline{\mathbf{r}}}$ by a channel matched filter (CMF) followed by a noise whitening filter in the Forney method \cite{Ungerboeck}. The complexity of this method including a whitening filter can be high, thus an alternative method based on Ungerboeck observation model can be choosen \cite{Ungerboeck}. In the Ungerboeck observation model, the minimum distance metric is constructed directly from the unwhitened CMF output, namely ${\mathbf{v}}\triangleq\underline{\underline{\mathbf{H}}}^H\underline{\underline{\mathbf{r}}}$. Taking ${{\mathbf{v}}}$ as the observation vector, the metric in (\ref{eqn:min_dist_metric_contd}) can be rewritten as
\begin{equation}
\label{eqn:min_dist_metric_contd_2}
\Lambda\left(\underline{\underline{\mathbf{r}}},\underline{\underline{\mathbf{H}}},\underline{\mathbf{x}}\right)=\gamma_2\mathrm{exp}\left(2\Re\left({{\mathbf{v}}}^H\underline{\mathbf{x}}\right)-\underline{\mathbf{x}}^H{{\mathbf{G}}}\hspace{2pt}\underline{\mathbf{x}}\right),
\end{equation}
where $\mathbf{G}\triangleq\underline{\underline{\mathbf{H}}}^H\underline{\underline{\mathbf{H}}}\triangleq\mathrm{blkToeplitz}(\underline{\mathbf{G}}^c,\underline{\mathbf{G}}^r)$, in which $\mathbf{G}^r=\left[ \mathbf{G}[0] \ \mathbf{G}[1] \ \cdots \ \mathbf{G}[L-1] \ \mathbf{0} \ \cdots \ \mathbf{0}\right]$, $\mathbf{G}^c=\left(\mathbf{G}^r\right)^H,$ where $\mathbf{G}[\ell]\triangleq\dfrac{a^2}{e+a^2N_o}\sum_{k=0}^{L-1-\ell}{{\mathbf{\tilde{H}}}}[k+\ell]^H{{\mathbf{\tilde{H}}}}[k].$

\begin{figure}[htbp]
\centering
\includegraphics[width=0.59\columnwidth]{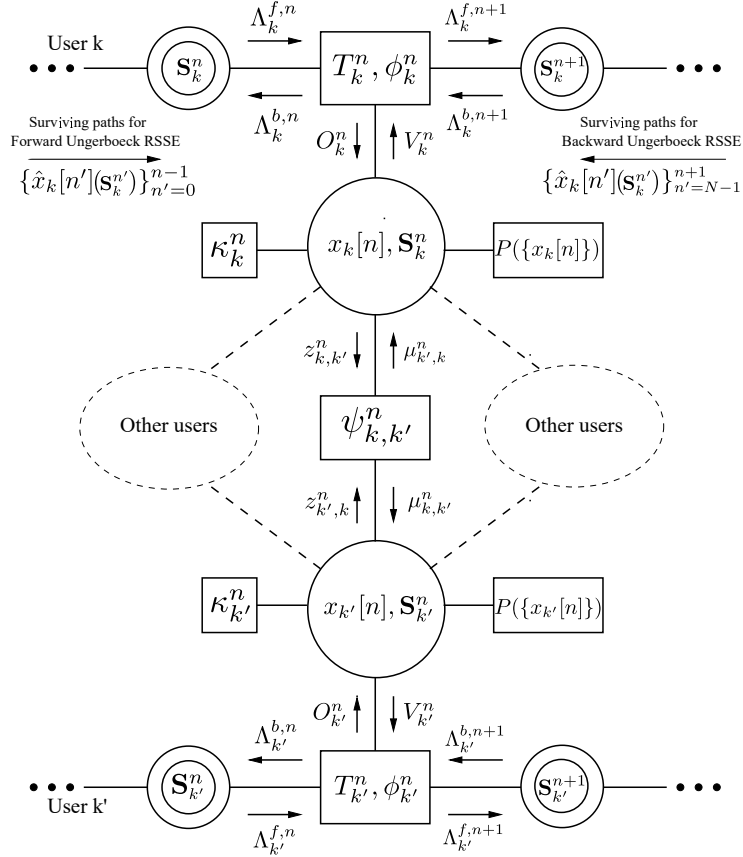}
\caption{Proposed factor graph corresponding to the calculation of the metric in (\ref{eqn:MU_prod_metric}).}
\label{fig:FG}
\end{figure}

With these definitions, the minimum distance metric in (\ref{eqn:min_dist_metric_contd_2}) can be computed recursively as
\begin{align}
\mathrm{ln}(\Lambda_N) &= \sum_{n=0}^{N-1}\left(\mathrm{ln}(\Lambda_{n+1})-\mathrm{ln}(\Lambda_n\right)) \nonumber \\
&= \sum_{n=0}^{N-1}\left(\sum_{k=1}^{K}\left[\kappa_k^n(v_k[n],x_k[n])
-\phi_k^n(x_k[n],\mathbf{S}_k^n)-\sum_{k'=1,k'<k}^K \psi_{k,k'}^{n}(x_k[n],\mathbf{S}_k^n,x_{k'}[n],\mathbf{S}_{k'}^n)\right]\right),
\label{eqn:MultiUserML_metric}
\end{align}
where $\Lambda_N=\Lambda\left(\underline{\underline{{v}}},\underline{\underline{\mathbf{H}}},\underline{\mathbf{x}}\right)$, $v_k[n]$ is the  $(Kn+k)^{th}$ element of $\mathbf{v}$, and $\mathbf{S}_k^n$ is the state vector of user $k$ at the $n^{th}$ time instant, which can be expressed as
\begin{equation}
\mathbf{S}_k^n = \left[x_{k}[n-1] \ \cdots \ x_{k}[n-J]\right].
\label{eqn:MU_State_Def_S_k}
\end{equation}
As can be noted in (\ref{eqn:MU_State_Def_S_k}), although we need to have $L-1$ elements in $\mathbf{S}_k^n$ for optimal sequence estimation, the number of elements in the state vector in (\ref{eqn:MU_State_Def_S_k}), namely $J$, can be selected to be less than $L-1$, to reduce the complexity of the detector. This can be done by utilizing surviving paths constructed based on the proposed Ungerboeck type reduced state sequence estimation (U-RSSE) with bidirectional decision feedback algorithm for MIMO, the details of which will be provided in the sequel. The functions $\kappa_k^n(.)$, $\phi_k^n$ and $\psi_{k,k'}^{n}(.)$ in (\ref{eqn:MultiUserML_metric}) are also defined as
\begin{align}
\kappa_k^n(.) &\triangleq 2\Re\left\{(v_k^*[n])x_k[n]\right\}-x_k^*[n][\mathbf{G}[0]]_{(k,k)}x_k[n],
\label{eqn:MultiUserML_metric_par_def_f} \\
\phi_k^n(.) &\triangleq 2\Re\left\{\zeta_{k,k}[n]\right\}, \psi_{k,k'}^{n}(.) \triangleq 2\Re\left\{x_{k'}^*[n][\mathbf{G}[0]]_{(k',k)}x_{k}[n]+\zeta_{k,k'}[n]+\zeta_{k',k}[n]\right\},
\label{eqn:MultiUserML_metric_par_def_h}
\end{align}
where $\zeta_{k,k'}[n]=\sum_{l=1}^{\mathrm{min}(L-1,n)}x_k^*[n][\mathbf{G}[\ell]]_{(k,k')}^Hx_{k'}[n-\ell]$. Here, $\kappa_k^n(.)$ can be regarded as the CMF output, $\phi_k^n(.)$ calculates the self-interference due to ISI, while $\psi_{k,k'}^{n}(.)$ corresponds to the interference caused to user $k$ by the other users. Employing (\ref{eqn:MultiUserML_metric})-(\ref{eqn:MultiUserML_metric_par_def_h}), a metric taking into account the \textit{a priori} probabilities of the transmitted data symbols can be found as
\begin{align} \mathrm{ln}\bigg(\Lambda\bigg(\left\{x_k[n],\mathbf{S}_k^n,\right.&\left.{v}_k^n\right\}_{\forall k,n}  \bigg)\bigg) 
 \nonumber \\ \propto\sum_{n=0}^{N-1} \sum_{k=1}^K \Bigg\{&\mathrm{ln}\left(P\left(\mathbf{S}_k^0\right)\right)+
\kappa_k^n\left(v_k^n,x_k[n]\right)-\phi_k^n\left(x_k[n],\mathbf{S}_k^n\right)
+\mathrm{ln}\left(T_k^n\left(x_k[n],\mathbf{S}_k^{n},\mathbf{S}_{k}^{n+1}\right)\right) \nonumber \\
&+\mathrm{ln}\left(P \left( \left\{x_k[n]\right\} \right)\right)- \sum_{k'=1,k'<k}^K \psi_{k,k'}^{n}\left(x_k[n],\mathbf{S}_k^n,x_{k'}[n],\mathbf{S}_{k'}^n\right)\Bigg\},
\label{eqn:MU_prod_metric}
\end{align}
where $P(\{x_k[n]\})$ and $P\left(\mathbf{S}_k^0\right)$ are the \textit{a priori} probabilities of the data symbol $x_k[n]$ and the initial state vector $\mathbf{S}_k^0$. Moreover, $\mathrm{ln}(.)$ takes the natural logarithm, and $T_k^n\left(x_k[n],\mathbf{S}_k^{n},\mathbf{S}_{k}^{n+1}\right)$ is the trellis indicator function, which is equal to 1 if a transition from $\mathbf{S}_k^{n}$ to $\mathbf{S}_k^{n+1}$ is possible with the data symbol being $x_k[n]$. Otherwise, it is equal to zero.
The proposed factor graph (FG) constructed for the calculation of (\ref{eqn:MU_prod_metric}) is presented in Fig.~\ref{fig:FG}. As can be noted in Fig.~\ref{fig:FG}, there are cycles of length 6. Although the existence of cycles in the FG in Fig.~\ref{fig:FG} result in approximate computation of  \textit{a posteriori} probabilities (APP) of each transmitted symbol, the approximation errors due to cycles are known to be negligable if the length of the cycles are greater than $4$ \cite{FG_cycle_length}. As can also be noted in Fig.~\ref{fig:FG}, the state vector $\mathbf{S}_k^n$ and the data symbol $x_k[n]$ are merged into a single variable node in order to increase the cycle length, which is known as streching in the literature \cite{factor_graph}.
Based on the FG in Fig.~\ref{fig:FG}, a novel reduced complexity quantization-aware Ungerboeck type message passing algorithm with bidirectional decision feedback (QA-UMPA-BDF) detector is proposed, which is characterized by the following message update rules based on sum-product algorithm (SPA) framework:
\begin{align}
&\Lambda_{k}^{f,n+1}\left(\mathbf{S}_{k}^{n+1}\right)=\mathrm{ln}\Bigg(\sum_{\thicksim\left\{\mathbf{S}_{k}^{n+1}\right\}}\mathrm{exp}\left(\Lambda_{k}^{f,n}\left(\mathbf{S}_k^n\right)+
\mathrm{ln}(T_k^n\left(.\right))-\phi_k^n\left(.\right) +V_k^n\left(x_k[n],\mathbf{S}_k^n\right)\right)\Bigg), \label{eqn:FG_messages_MUD_lambda_f}
\end{align}
\begin{align}
&\Lambda_{k}^{b,n}\left(\mathbf{S}_k^n\right)=\mathrm{ln}\Bigg(\sum_{\thicksim\left\{\mathbf{S}_k^n\right\}}\mathrm{exp}\left(\Lambda_{k}^{b,n+1}\left(\mathbf{S}_{k}^{n+1}\right)+
\mathrm{ln}\left(T_k^n\left(.\right)\right)-\phi_k^n\left(.\right) +V_k^n\left(x_k[n],\mathbf{S}_k^n\right)\right)\Bigg), \label{eqn:FG_messages_MUD_lambda_b}
\end{align}
\vspace{-20pt}
\begin{align}
O_k^n\left(x_k[n],\mathbf{S}_k^n\right)&=\Lambda_{k}^{f,n}\left(\mathbf{S}_k^n\right)+\Lambda_{k}^{b,n+1}\left(\mathbf{S}_{k}^{n+1}\right)
+\mathrm{ln}\left(T_k^n\left(.\right)\right)-\phi_k^n\left(.\right), \label{eqn:FG_messages_MUD_lambda_V}\\
V_k^n\left(x_k[n],\mathbf{S}_k^n\right)&=\mathrm{ln}\left(P \left( \left\{x_k[n]\right\} \right)\right) +\kappa_k^n\left(.\right)+\left[\sum_{l=1,\;l \neq k}^K \mu_{l,k}^n\left(x_k[n],\mathbf{S}_k^n\right)\right], \label{eqn:FG_messages_MUD_O}\\
\mu_{k',k}^{n}\left(x_k[n],\mathbf{S}_k^n\right)&=\mathrm{ln}\Bigg(\sum_{\left\{x_{k'}[n],\mathbf{S}_{k'}^n\right\}} \exp\left(z_{k',k}^{n}\left(x_{k'}[n],\mathbf{S}_{k'}^n\right)-\psi_{k,k'}^{n}\left(.\right)\right)\Bigg),  \label{eqn:FG_messages_MUD_nu} \\
z_{k,k'}^{n}\left(x_k[n],\mathbf{S}_k^n\right)&={O_k^n\left(x_k[n],\mathbf{S}_k^n\right)+V_k^n\left(x_k[n],\mathbf{S}_k^n\right)}-
{\mu_{k',k}^{n}\left(x_k[n],\mathbf{S}_k^n\right)},
\label{eqn:FG_messages_MUD_mu}
\end{align}
where $\sum_{\thicksim\left\{x \right\}}$ is defined as the sum over all variables excluding $x$. Note that we calculate messages in log-domain to avoid numerical issues due to large numbers as multiplications performed in SPA are reflected as summations in log domain in (\ref{eqn:FG_messages_MUD_lambda_f})-(\ref{eqn:FG_messages_MUD_mu}). For further avoidance of numerical issues, the max-log approximation \cite{max_log} can be used for (\ref{eqn:FG_messages_MUD_lambda_f}), (\ref{eqn:FG_messages_MUD_lambda_b}) and (\ref{eqn:FG_messages_MUD_nu}) as
\begin{align}
\label{eqn:FG_messages_MUD_lambda_f_approx}
\Lambda_{k}^{f,n+1}\left(\mathbf{S}_{k}^{n+1}\right)&\approx\max_{\left\{\mathbf{S}_{k}^n\right\}}\left(\Lambda_{k}^{f,n}\left(\mathbf{S}_k^n\right)+
\mathrm{ln}(T_k^n\left(.\right))-\phi_k^n\left(.\right) +V_k^n\left(x_k[n],\mathbf{S}_k^n\right)\right),\\
\label{eqn:FG_messages_MUD_lambda_b_approx}
\Lambda_{k}^{b,n}\left(\mathbf{S}_{k}^{n}\right)&\approx\max_{\left\{\mathbf{S}_{k}^{n+1}\right\}}\left(\Lambda_{k}^{b,n+1}\left(\mathbf{S}_k^{n+1}\right)+
\mathrm{ln}(T_k^n\left(.\right))-\phi_k^n\left(.\right) +V_k^n\left(x_k[n],\mathbf{S}_k^n\right)\right),\\
\label{eqn:FG_messages_MUD_nu_approx}
\mu_{k',k}^{n}\left(x_k[n],\mathbf{S}_k^n\right)&\approx\max_{\left\{x_{k'}[n],\mathbf{S}_{k'}^n\right\}} \left( z_{k',k}^{n}\left(x_{k'}[n],\mathbf{S}_{k'}^n\right) -\psi_{k,k'}^{n}\left(.\right)\right).
\end{align}
Regarding the interpretation of the FG in Fig.~\ref{fig:FG}, the upper part is responsible for the RSSE operation with the help of bidirectional decision feedback providing the surviving paths needed for metric calculations. The message $\mu_{k',k}^{n}$ in (\ref{eqn:FG_messages_MUD_nu}) is responsible for soft multi-user interference (MUI) cancellation between user $k$ and user $k'$, whereas $V_k^n$ contains the information for the interference of all users to user $k$ and the CMF output for the $n^{th}$ symbol of user $k$.

\subsection{Bias Compensation}
Owing to the state reduction and the pre-cursor ISI remaining after CMF operation, an anti-causal interference appears. As a result, U-RSSE suffers from $\textrm{\it{correct path loss}}$ even when there is no noise and multi-user interference, as pointed in our previous work for unquantized single-input single-output (SISO) systems \cite{gmguvensen_2014_reduced_state}. This interference results in a \textit{bias} affecting the tentative decisions in a survivor map. This bias has to be corrected in the forward surviving path construction in a similar manner as performed in \cite{gmguvensen_2014_reduced_state}. With such a correction, the surviving path construction of the states of the $k^{th}$ user can be found as
\begin{equation}
\label{eqn:surv_path_wh_bias_corr}
\hat{x}_{k}[n-{J}](\mathbf{S}_k^{n})=\underset{{x}_{k}[n-{J}]}{\operatorname{arg\;max}} \left[ \Lambda_k^{f,n}\left(S_k^n\right) + \phi_k^n\left(.\right) + V_k^n\left(.\right)-\beta_{k}^{n-{J}}\left(\mathbf{S}_k^n,{x}_{k}[n-{J}]\right)\right],
\end{equation}
where $\beta_{k}^{n-{J}}\left(.\right)$ is the \textit{bias} correction term. The bias correction term can be found by replacing the $\mathbf{h}_n^H\mathbf{G}_{\hat{I}_{\ell_1}^m(\mathbf{S}_k^m),\tilde{I}_{\ell_1+\ell_2}^{n}}^{(m,n)}(\ell_2)\mathbf{h}_m$ terms in \cite[Eqn. (4.27)]{{guvensen2014reduced}}, which is our previous work finding the \textit{bias} term for a U-RSSE receiver for a multiple-input single-output (MISO) scenario, with ${x}^*_k[l_1](\mathbf{S}_k^n)[\mathbf{G}[l_2]]_{(k,k')}\tilde{x}_{k'}[l_1+l_2]$ in this work. In that case, $\beta_{k}^{n-{J}}\left(\mathbf{S}_k^n,{x}_{k}[n-{J}]\right)$ is found as
\begin{align}
\hspace{-8pt}
\beta_{k}^{n-J}\left(\mathbf{S}_k^n,{x}_{k}[n-{J}]\right)= 2\operatorname{Re}&\left\{\sum_{k'=1}^K\left[\sum_{l_1=n-L+2}^{n-J}\;\sum_{l_2=n-l_1+1}^{L-1} 
{x}^*_k[l_1](\mathbf{S}_k^n)[\mathbf{G}[l_2]]_{(k,k')}\tilde{x}_{k'}[l_1+l_2]\right]\right\}
\label{eqn:bias_correction_MUD}
\end{align}
where ${x}^*_k[l_1](\mathbf{S}_k^n)$ for $l_1<n-j$ can be found from the surviving paths constructed using (\ref{eqn:surv_path_wh_bias_corr}) at the previous time instants and $\tilde{x}_{k'}[l_1+l_2]$ can also be found from the hard tentative decisions about future symbols, obtained in the previous iterations (what is meant by ``iterations" will be detailed in Section~\ref{sec:mess_pass_schedule}). The bias term is also simplified for the full decision feedback case (when no state is used, that is, when $J=0$) as
\begin{align}
\beta_{k}^{n}\left({x}_{k}[n]\right)= 2\operatorname{Re}&\left\{\sum_{k'=1}^K\left[\sum_{l_2=1}^{L-1} 
{x}^*_k[n][\mathbf{G}[l_2]]_{(k,k')}\tilde{x}_{k'}[n+l_2]\right]\right\},
\label{eqn:bias_correction_MUD_DFB}
\end{align}
since the terms of the outer summation with index $\ell_1\neq n-J=n$ in (\ref{eqn:bias_correction_MUD}) can be omitted as the maximization is over $x_k[n]$ in (\ref{eqn:surv_path_wh_bias_corr}) for $J=0$. Ultimately, the marginalized version of the metric in (\ref{eqn:MU_prod_metric}) can be calculated in the termination step of the SPA as
\begin{align}
\mathrm{ln}\bigg(\Lambda\bigg( x_k[n],\mathbf{S}_k^n, \left\{{v}_k^n\right\}_{\forall k,n} \bigg)\bigg)&=
\sum_{S_k^n}\left[V_k^n\left(.\right)+  
O_k^n\left(.\right)-\lambda_{k}^{n}\left({x}_{k}[n],\mathbf{S}_k^n\right)\right],
\label{eqn:FG_MAP_rule_MUD}\\
\lambda_{k}^{n}\left({x}_{k}[n],\mathbf{S}_k^n\right)= 2\operatorname{Re}\left\{\sum_{k'=1}^K\left[\sum_{l_1=n-L+J+2}^{n}\right.\right.&\left.\left.\sum_{l_2=n-l_1+J+1}^{L-1} 
{x}^*_k[l_1](\mathbf{S}_k^n)[\mathbf{G}[l_2]]_{(k,k')}\tilde{x}_{k'}[l_1+l_2]\right]\right\} 
\label{eqn:bias_final}
\end{align}
with the corresponding data symbol estimates maximizing the metric in (\ref{eqn:MU_prod_metric}) given as
\begin{equation}
\hat{x}_k[n]=\underset{\{x_k[n]\}}{\operatorname{arg\;max}}\;
\sum_{\mathbf{S}_k^n}\left[V_k^n\left(x_k[n],\mathbf{S}_k^n\right)  
+O_k^n\left(x_k[n],\mathbf{S}_k^n\right)-\beta_{k}^n\left(x_k[n],\mathbf{S}_k^n\right)\right].
\label{eqn:MAP_decision_MUD}
\end{equation}

\subsection{Message Passing Schedule}
\label{sec:mess_pass_schedule}
Owing to the cycles existing in the FG in Fig.~\ref{fig:FG}, there is no unique message passing schedule for SPA operation. Therefore, we employ a serial schedule as in \cite{factor_graph} for updating the messages. The proposed scheduling for forward recursion in time-domain is presented in Algorithm 1.
\begin{algorithm}
  \caption{QA-UMPA-BDF, forward recursion in time-domain}\label{qa_umpa_bdf_algorithm}
  \hspace*{\algorithmicindent} \textbf{Input}: MF output $\mathbf{v}$ and correlation metric $\mathbf{G}$. \\
  \hspace*{\algorithmicindent} \textbf{Initialization}: Initialize all messages $z_{k',k}^n$, $\mu_{k',k}^n$, $V_k^n$, $O_k^n$, $\Lambda_k^{f,n}$, $\Lambda_k^{b,n}$ as zero. 
  \begin{algorithmic}[1]
      \For{$n=0:1:N-1$}
      \For{$k=1:1:K$}
      \State{ \texttt{Update $O_{k}^{n}\left(.\right)$ using (\ref{eqn:FG_messages_MUD_lambda_V}).}}
      \EndFor
      \State{Forward recursion in user domain:}
      \For{$k=1:1:K$}
      \For{$k'=1:1:k-1$}
        \State \texttt{Update $\mu_{k',k}^n$ using (\ref{eqn:FG_messages_MUD_nu_approx}).}
        \EndFor
        \State \texttt{Update the term $V_k^n\left(.\right)$ using (\ref{eqn:FG_messages_MUD_O}).}
        \For{$k'=k+1:1:K$}
		\State \texttt{Update $z_{k,k'}^{n}$ using (\ref{eqn:FG_messages_MUD_mu}).}
      \EndFor
      \EndFor
      \State{Backward recursion in user domain:}
      \For{$k=K:-1:1$}
      \For{$k'=K:-1:k+1$}
        \State \texttt{Update $\mu_{k',k}^{n}$ using (\ref{eqn:FG_messages_MUD_nu_approx}).}
        \EndFor
        \State \texttt{Update the term $V_{k}^n\left(.\right)$ using (\ref{eqn:FG_messages_MUD_O}).}
        \For{$k'=k-1:-1:1$}
		\State \texttt{Update $z_{k,k'}^{n}$ using (\ref{eqn:FG_messages_MUD_mu}).}
      \EndFor
      \EndFor
      \State{Update the time-domain forward messages:}
      \For{$k=1:1:K$}
      \State \texttt{Update $\Lambda_{k}^{f,n+1}\left(.\right)$ using (\ref{eqn:FG_messages_MUD_lambda_f_approx}).}
   	  \State \texttt{Calculate bias term $\beta_{k}^{n-J}\left(\mathbf{S}_k^n,{x}_{k}[n-{J}]\right)$ using (\ref{eqn:bias_correction_MUD}) or (\ref{eqn:bias_correction_MUD_DFB}).}
      \State \texttt{Update surviving paths $\hat{x}_{k}[n-{J}](\mathbf{S}_k^{n})$ using (\ref{eqn:surv_path_wh_bias_corr}).}
      \EndFor
      \EndFor
  \end{algorithmic}
\end{algorithm}

%
%
%
%

When the forward recursion in time-domain in Algorithm 1 ends, the same procedure is performed as the backward recursion in time-domain, except that the time index at the outermost for-loop in Algorithm 1 will be from $N-1$ to $0$, the operation in line 27 will be replaced by an update of $\Lambda_k^{b,n}$ using (\ref{eqn:FG_messages_MUD_lambda_b_approx}), and the lines 28-29 will not be performed. Completion of forward and backward recursions in time-domain constitutes an iteration of QA-UMPA-BDF. Although various choices can be made for stopping criteria, the one adopted in this study is the completion of a predefined number of iterations. The initialization step in Algorithm 1 should only be performed for the forward recursion in time-domain at the first iteration.


\subsection{Computational Complexity Analysis}
The computational complexity per iteration of the proposed QA-UMPA-BDF detector can be found by analyzing (\ref{eqn:FG_messages_MUD_lambda_f})-(\ref{eqn:bias_correction_MUD}). For the complexity analysis we consider the max-log approximations for (\ref{eqn:FG_messages_MUD_lambda_f}), (\ref{eqn:FG_messages_MUD_lambda_b}) and (\ref{eqn:FG_messages_MUD_nu}), which are (\ref{eqn:FG_messages_MUD_lambda_f_approx}), (\ref{eqn:FG_messages_MUD_lambda_b_approx}), (\ref{eqn:FG_messages_MUD_nu_approx}). The complexity (number of flops) to calculate the messages per single iteration of the proposed detector is provided in Table~\ref{tbl:comput_comp}.
\begin{table}[htbp]
\centering
\begin{tabular}{|c|c|c|c|c|c|c|}
\hline
    & (\ref{eqn:FG_messages_MUD_lambda_V}) & (\ref{eqn:FG_messages_MUD_O}), (\ref{eqn:FG_messages_MUD_mu}) & (\ref{eqn:FG_messages_MUD_lambda_f_approx}), (\ref{eqn:FG_messages_MUD_lambda_b_approx}) & (\ref{eqn:FG_messages_MUD_nu_approx}) &  (\ref{eqn:surv_path_wh_bias_corr}), (\ref{eqn:bias_correction_MUD})   \\ \hline
Complexity & $\mathcal{O}\left(NP^{(J+1)}KL\right)$  & $\mathcal{O}\left(NP^{(J+1)}K^2\right)$   & $\mathcal{O}\left(NP^{(J+1)}K\right)$   & $\mathcal{O}\left(NP^{2(J+1)}K\right)$ & $\mathcal{O}\left(NP^{(J+1)}K^2L\right)$ \\ \hline
\end{tabular}
\caption{Computational complexity of the QA-UMPA-BDF detector per iteration. \label{tbl:comput_comp}}
\end{table}
As can be noted, the computational complexity per iteration can be as high as $\mathcal{O}\left(NP^{2(L+1)}K\right)$ if reduced state estimation is not employed (when $J=L-1$). However, the computational complexity can be reduced to $\mathcal{O}\left(NPK^2\right)+\mathcal{O}\left(NP^{2}K\right)+\mathcal{O}\left(NPK^2L\right)$ for $J=0$, which changes linearly with $N$, $L$, and quadratically with $K$ and $P$. The complexity to calculate CMF output ${\mathbf{v}}$ and the correlation metric $\mathbf{G}$ are $\mathcal{O}\left(NMKL\right)$ and $\mathcal{O}\left(MK^2L\right)$, which are excluded from discussion as they are only calculated once, not per iteration. In the representative benchmark algorithm that we compare the proposed QA-UMPA-BDF detector, namely the ``Robust MMSE" in \cite[Eqn.(27)]{dinnis_benchmark_sc_fde}, the computational complexity is $\mathcal{O}\left(MKN\mathrm{log}_2(N)\right)+\mathcal{O}\left(NMK\right)+\mathcal{O}\left(NMK^2\right)+\mathcal{O}\left(NK^3\right)+\mathcal{O}(NKP)$, whose complexity is growing with $K^3$. Therefore, the proposed QA-UMPA-BDF detector for $J=0$ has lower complexity compared to the benchmark detector, especially when $K$ is large. We will also show in Section~\ref{sec:sim_res} that the proposed detector can converge in about $I=2$ iterations for most of the cases. Therefore, the number of iterations does not increase the proposed detector complexity to a significant degree. Lower complexity detectors compared to the ``Robust MMSE" detector are also proposed in \cite{dinnis_benchmark_sc_fde}. However, their performance is inferior compared to ``Robust MMSE" detector \cite[Fig.12]{dinnis_benchmark_sc_fde}, thus ``Robust MMSE" is chosen as the benchmark detector.
\section{Performance Metrics}
To assess the performance of the proposed LMMSE channel estimator, normalized MSE (nMSE) will be used as a metric. The nMSE taking into account the channel coefficients multiplied by the power-delay profile can be found as \cite{nmse}
\begin{equation}
\label{eqn:nmse}
\mathrm{nMSE}=\dfrac{\mathrm{Tr}\left[\mathbf{\Omega}\mathbf{C}_{\underline{\mathbf{\hat{h}}}}^{\text{LMMSE}}\mathbf{\Omega}^H\right]}{\mathrm{Tr}\left[\mathbf{\Omega}\mathbf{C}_{\underline{\mathbf{{h}}}}\mathbf{\Omega}^H\right]}=\dfrac{{\mathrm{Tr}\left[\mathbf{\Omega}\mathbf{C}_{\underline{\mathbf{\hat{h}}}}^{\text{LMMSE}}\mathbf{\Omega}^H\right]}}{MK},
\end{equation}
where $\mathbf{\Omega}$ is a diagonal matrix whose $(ML(k-1)+M\ell+1)^{th}$ to $(ML(k-1)+M\ell+M)^{th}$ diagonal elements are all equal to $\sqrt{\rho_{k}[\ell]}$. $\mathbf{C}_{\underline{\mathbf{\hat{h}}}}^{\text{LMMSE}}$ can be found by (\ref{eqn:mse}), ({\ref{eqn:mse_white_noise}) or (\ref{eqn:LMMSE_white_noise_2}).
For the data detector performance metric, we use uncoded bit-error-rate (BER) and average mismatched achievable rate (AIR) per user \cite{mismatched_dec_lapi}, which is a suitable metric to assess the performance of mismatched detectors employing approximate APPs for detection, as the exact APPs cannot be calculated due to the cycles in the FG in Fig.~\ref{fig:FG} and Gaussian effective noise approximations. The mismatched average AIR per user can be expressed as \cite{mismatched_dec_lapi}
\begin{equation}
\label{eqn:AIR_def}
\mathrm{AIR}=\mathbb{E}_{\underline{\mathbf{x}},\hspace{1pt} \underline{\mathbf{H}}}\left[\frac{1}{NK}\sum_{k=1}^{K}\sum_{n=0}^{N-1}\left[\mathrm{log}_2(P)-\mathrm{log}_2\left(\dfrac{\sum_{{x'_k[n]}\in A_{{x}}}\tilde{p}(\mathbf{v}|x'_k[n])}{\tilde{p}(\mathbf{v}|\hat{x}_k[n]=x_k[n])}\right)\right]\right],
\end{equation}
where $A_{{x}}$ is the set of all possible constellation points, $x_k[n]$ is the correct value of the transmitted symbol, and $\tilde{p}(\mathbf{v}|x_k[n])$ are the approximate APPs which are found using (\ref{eqn:MAP_decision_MUD}) as
\begin{equation}
\tilde{p}(\mathbf{v}|x_k[n])\propto\sum_{\mathbf{S}_k^n}\exp\left(V_k^n\left(x_k[n],\mathbf{S}_k^n\right)  
+O_k^n\left(x_k[n],\mathbf{S}_k^n\right)-\beta_{k}^n\left(x_k[n],\mathbf{S}_k^n\right)\right).
\end{equation}

\section{Simulation Results}
\label{sec:sim_res}
In this section, we will present the nMSE, uncoded BER and AIR performance of the proposed LMMSE channel estimator and QA-UMPA-BDF detector. We will mostly concentrate on the performance comparison between the proposed QA-UMPA-BDF detector and the representative robust MMSE detector \cite{dinnis_benchmark_sc_fde} from the literature, which has comparable complexity to our detector, as the reduced state length of the QA-UMPA-BDF detector is set as $J=0$. We will see that the proposed detector outperforms the representative detector in all cases, even if their complexities are similar and the QA-UMPA-BDF detector provides a higher spectral efficiency, due to the absence of a cyclic-prefix. Unless otherwise stated, $M=100$, the power-delay profile of the transmission channel is COST-207 typical delay profile for suburban and urban areas \cite{proakis2007digital}. The number of channel taps $L=32$, with the power ratio of the first and the last taps being $30$ dB. The number of iterations for the QA-UMPA-BDF detector is selected as $2$. The pilot symbols are created as random complex numbers from QPSK modulation. $E_b\triangleq E_s/\mathrm{log}_2(P)$ corresponds to the bit-energy. LMMSE channel estimates and MSE values are found based on (\ref{eqn:LMMSE_white_noise}) and (\ref{eqn:mse_white_noise}). The step size of the quantizer is selected to optimally to minimize quantization noise as in \cite{chan_est_freq_sel_6}.

To determine the necessary training length for the channel estimation, the nMSE or BER vs. the training length ($\tau$) performances are obtained as in Fig.~\ref{fig:nmse_ber_training_length}.
\begin{figure}[htbp]
	\centering
	\begin{subfigure}{0.40\textwidth} 
		\includegraphics[width=\textwidth]{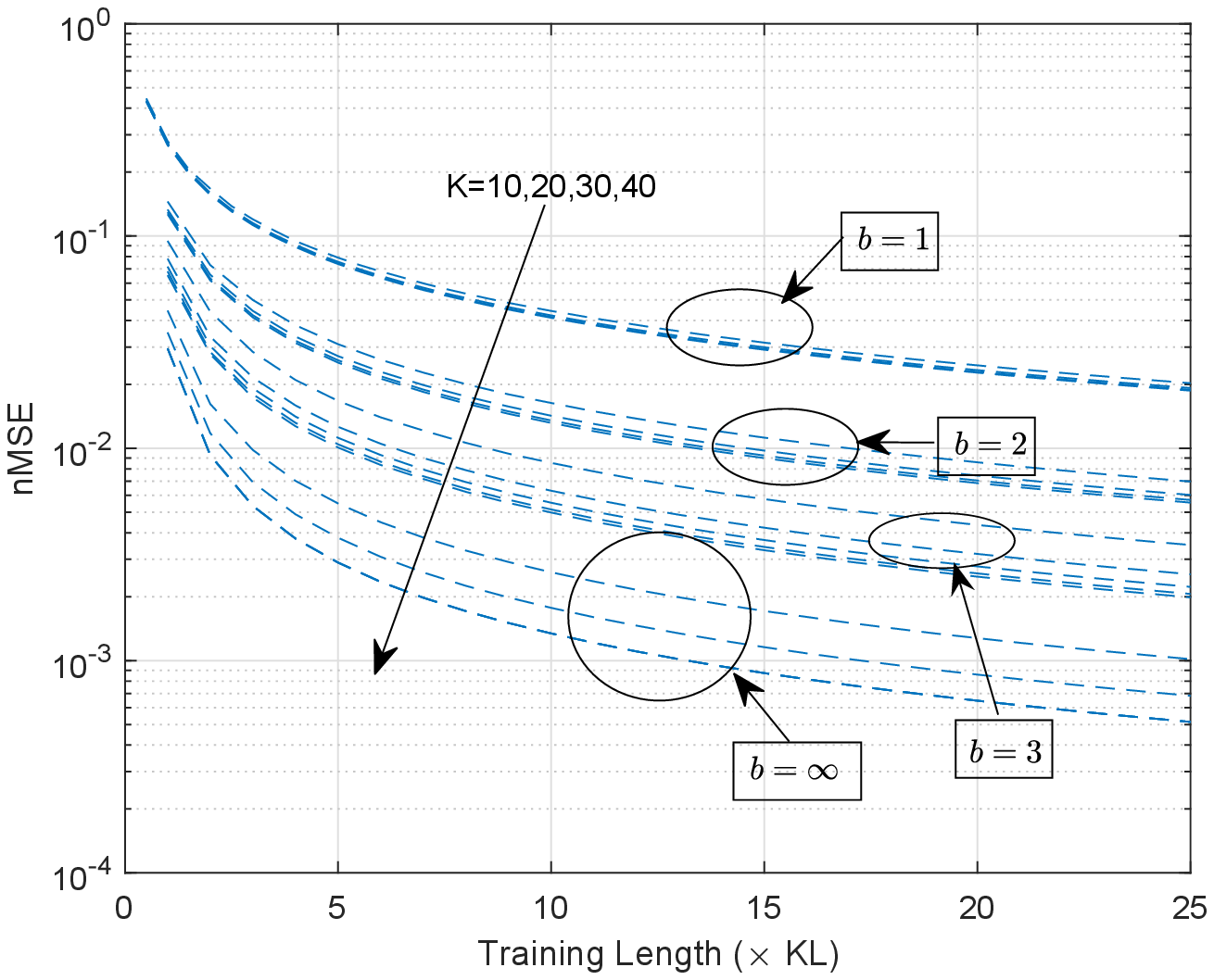}
		\caption{nMSE vs. training length ($\tau$), $E_b/N_o=0 \ \mathrm{dB}$, $K=10,20,30,40$.} 
		\label{fig:nmse_training_length}
	\end{subfigure}
	\hspace{1em} 
	\begin{subfigure}{0.40\textwidth} 
		\includegraphics[width=\textwidth]{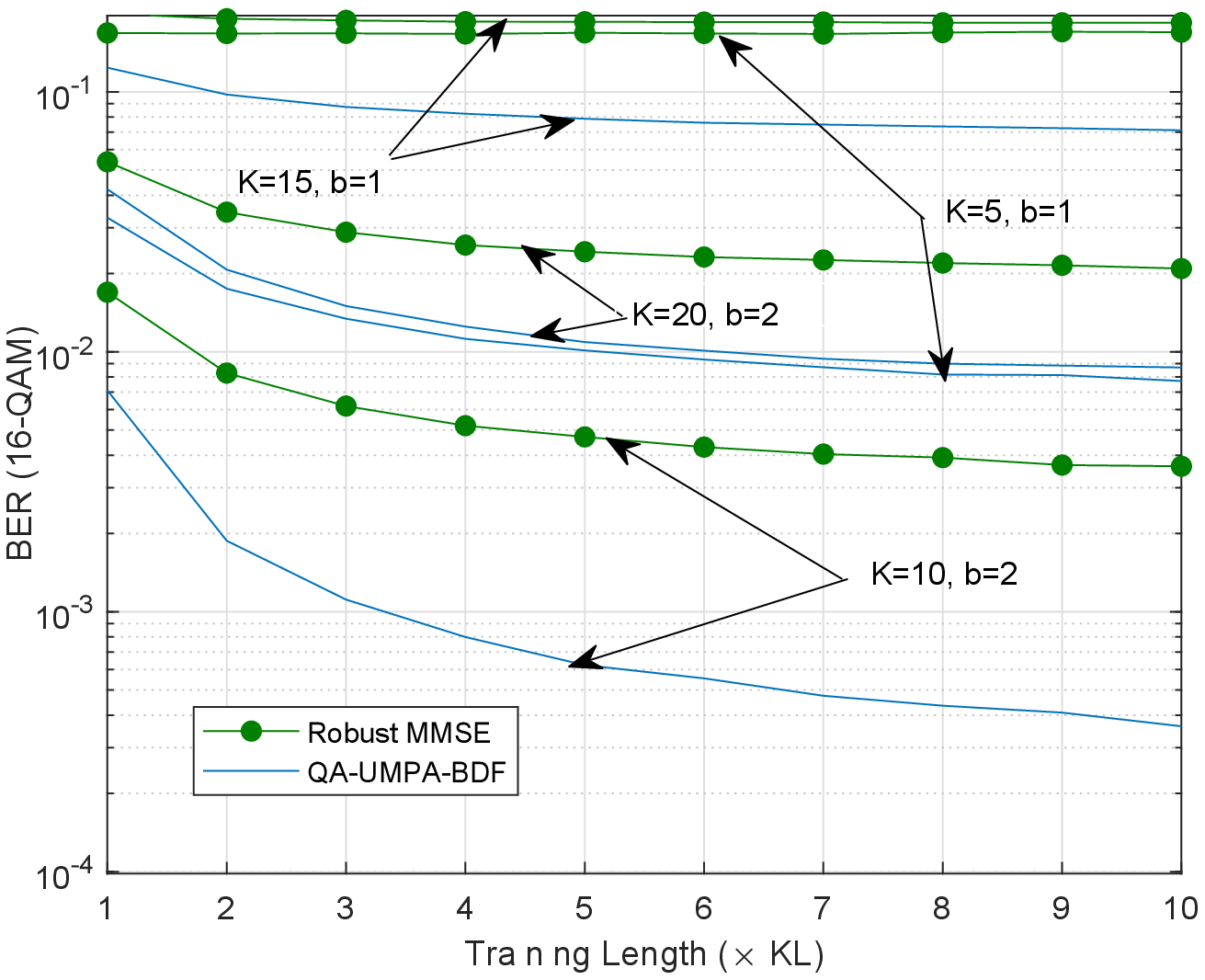}
		\caption{BER vs. training length, $K=5,10,15,20$, $E_b/N_o=0\ \mathrm{dB}$, 16-QAM.} 
		\label{fig:ber_training_length}
	\end{subfigure}
	\caption{nMSE (a) and BER (b) vs. training length ($\tau$).}
	\label{fig:nmse_ber_training_length} 
\end{figure}
In Fig.~\ref{fig:nmse_training_length}, we observe that if we need to have a nMSE level less than $10^{-1}$, $\tau\geq 5KL$ will be an adequate choice for the proposed LMMSE channel estimator for one-bit quantizer, although this number is much less for higher bit resolutions. However, we can also say that there is a decreased improvement for nMSE if the training length $\tau>5KL$ for any bit resolution. Nevertheless, observing nMSE alone may not be enough to foresee how the error-rate performance of the proposed detector changes with the training length. Therefore, BER vs. training length is also obtained for 16-QAM modulated data symbols as in Fig.~\ref{fig:ber_training_length}. As can be noted in Fig.~\ref{fig:ber_training_length}, for $1$ and $2$ bits, there is a significant error-floor advantage of the proposed detector compared to the Robust MMSE detector \cite{dinnis_benchmark_sc_fde} for all cases. Moreover, we can see that with very low resolution quantizers (1 or 2 bits), increasing $\tau$ more than $5KL$ is not very effective for decreasing BER. Therefore, we will set $\tau=5KL$ when $q=1,2$. For $q=3$, we will set $\tau=3KL$ observing nMSE values close to $10^{-2}$ from Fig.~\ref{fig:nmse_ber_training_length}, which is considered to be adequate. For larger bit resolutions, $\tau$ will be selected as $2KL$ in the subsequent simulations, all performed under imperfect channel state information (CSI).

In Fig.~\ref{fig:BER_SNR_QPSK_16QAM_b1_2}, we compare the BER performance of the proposed QA-UMPA-BDF and the  Robust MMSE \cite{dinnis_benchmark_sc_fde} detectors in Fig.~\ref{fig:BER_SNR_QPSK_16QAM_b1_2} for either QPSK with $q=1$ or 16-QAM with $q=2$. We also include the performance of a genie aided detector, which calculates the metric in (\ref{eqn:MU_prod_metric}) for an $x_k[n]$ assuming that all other symbols are perfectly known so that the terms corresponding to the ISI and MUI cancellation in (\ref{eqn:MU_prod_metric}) are calculated accordingly\footnote{Genie aided detector uses perfect bidirectional decision feedback while constructing surviving paths in (\ref{eqn:surv_path_wh_bias_corr}) and bias terms in (\ref{eqn:bias_correction_MUD}). For unquantized case and perfect CSI, the performance of this detector corresponds to matched filter bound \cite{gmguvensen_2014_reduced_state,guvensen2014reduced}.}. Genie aided detector performance is mainly limited by thermal and quantization noise and named ``Genie Aided Det." in all figures.

\begin{figure}[htbp]
	\centering
	\begin{subfigure}{0.41\textwidth} 
		\includegraphics[width=\textwidth]{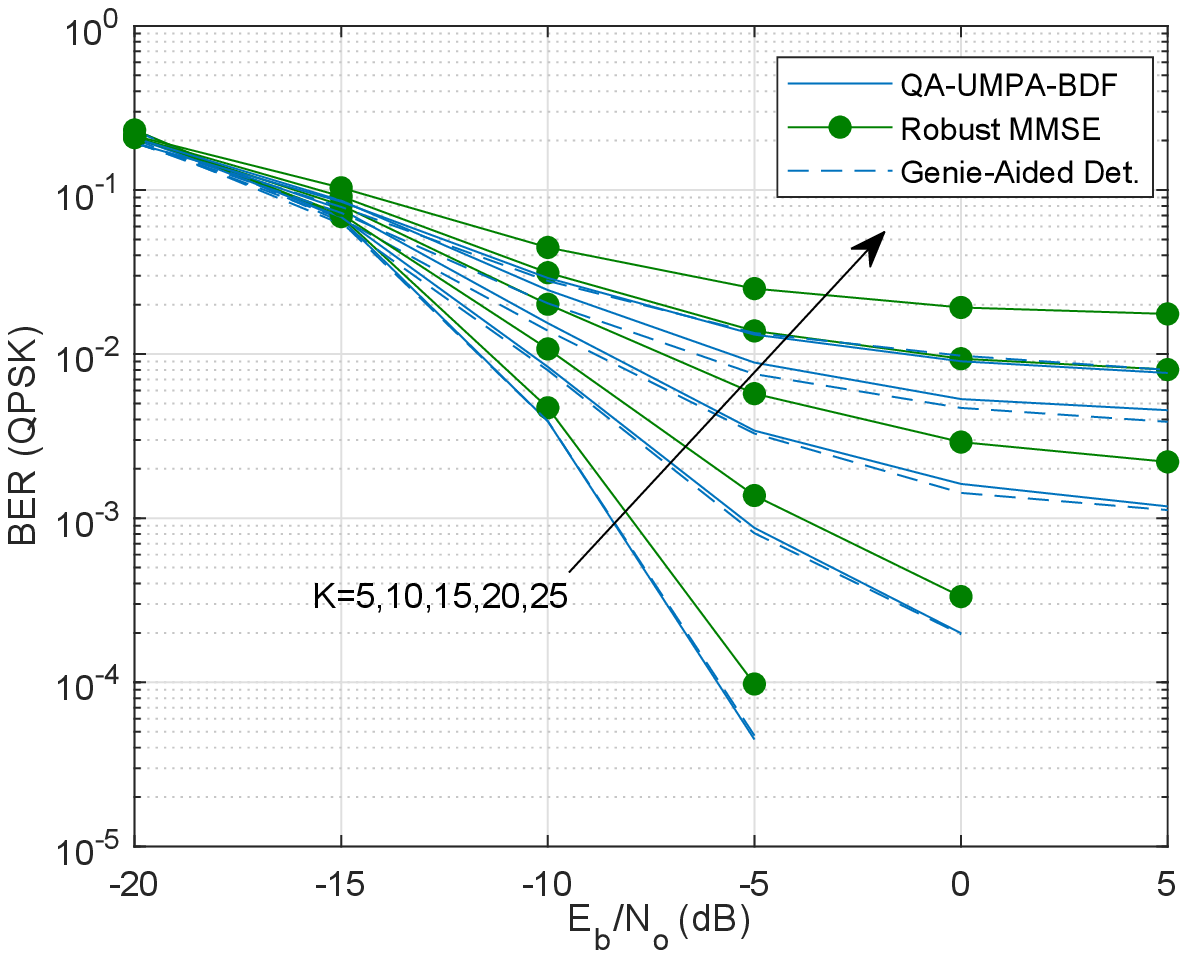}
		\caption{BER vs. $E_b/N_o$ for QPSK, $q=1$.} 
		\label{fig:BER_SNR_QPSK_b1}
	\end{subfigure}
	\hspace{1em} 
	\begin{subfigure}{0.41\textwidth} 
		\includegraphics[width=\textwidth]{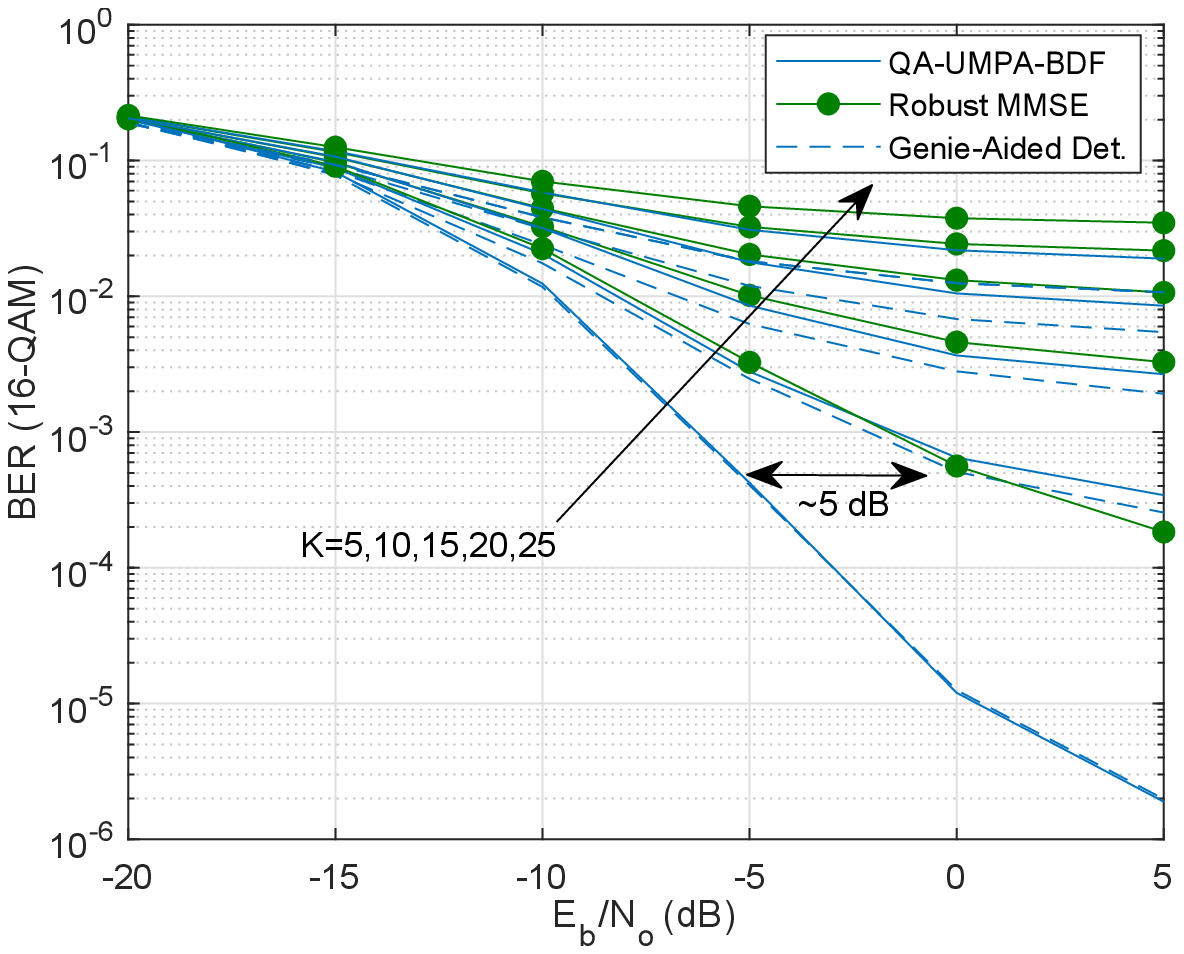}
		\caption{BER vs. $E_b/N_o$ for 16-QAM, $q=2$.} 
		\label{fig:BER_SNR_16QAM_b2}
	\end{subfigure}
	\caption{BER vs. $E_b/N_o$ for QPSK, $q=1$ (a) or 16-QAM, $q=2$ (b).}
	\label{fig:BER_SNR_QPSK_16QAM_b1_2} 
\end{figure}

As can be observed in Fig.~\ref{fig:BER_SNR_QPSK_b1}, the QA-UMPA-BDF detector has better performance compared to the representative benchmark detector for all number of user values (for $K=5,10,\ldots,25$), despite being spectrally more efficient due to the CP free transmission. If the modulation type is changed to 16-QAM, the SNR advantage of QA-UMPA-BDF is up to $5$ dB as can be noted in Fig.~\ref{fig:BER_SNR_16QAM_b2}. 
Moreover, the QA-UMPA-BDF performance is always very close to genie-aided detector performance with only $2$ iterations, which is the case in most of the subsequent simulations.

As the next simulation scenario, we plot the error-rate performances for fixed $K$ but varying $q$ in Fig.~\ref{fig:BER_SNR_QPSK_16QAM_all_b}. For all cases, again QA-UMPA-BDF has better performance, where the performance gap  between the two detectors is widened for 16-QAM. For QPSK, performance improvement is not much for $q>2$, whereas $q=3$ seems to be an enough for 16-QAM. Again, two iterations for QA-UMPA-BDF is observed to be sufficient to attain genie-aided detector performance.
\begin{figure}[htbp]
	\centering
	\begin{subfigure}{0.39\textwidth} 
		\includegraphics[width=\textwidth]{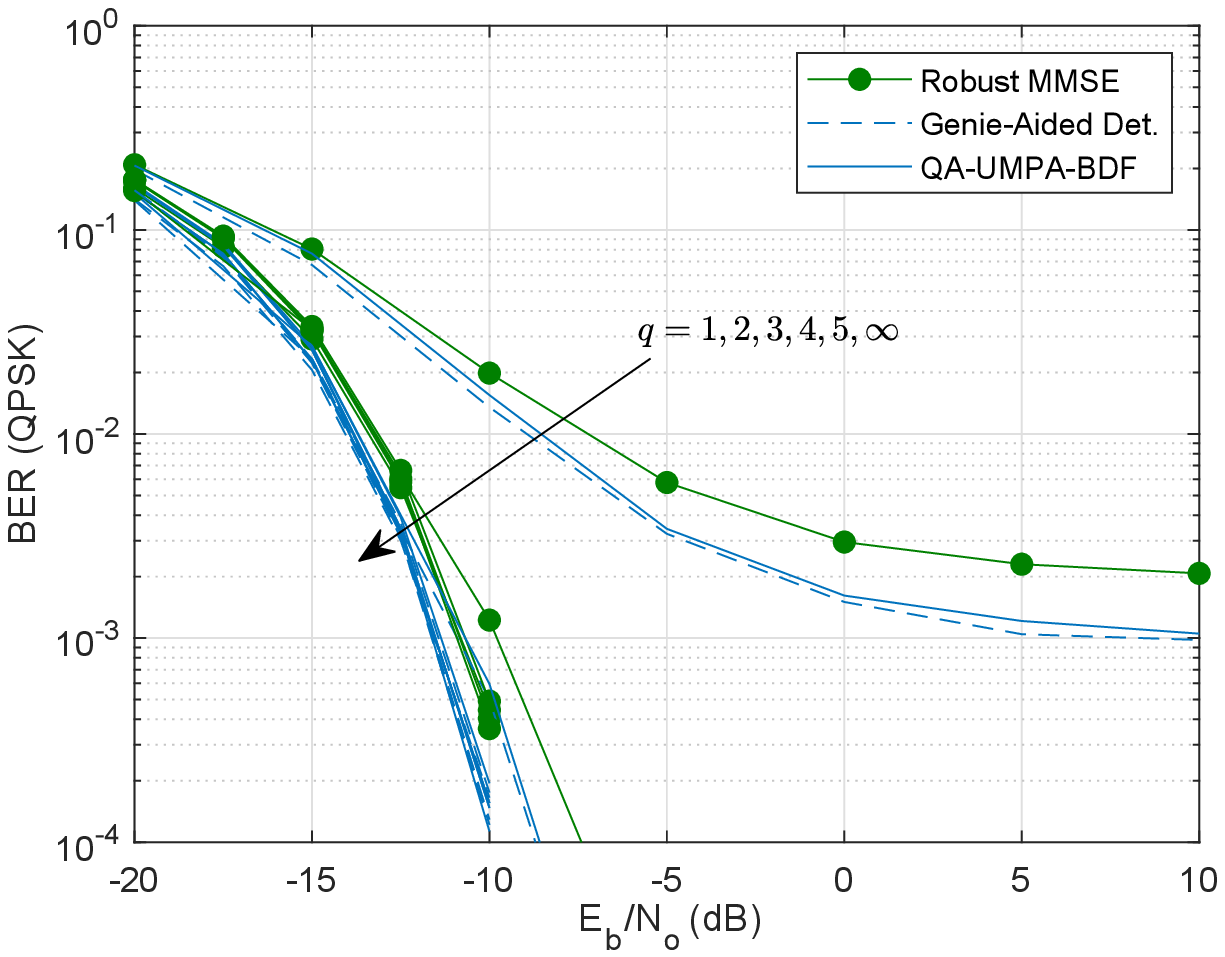}
		\caption{QPSK.} 
		\label{fig:BER_SNR_QPSK_all_b}
	\end{subfigure}
	\hspace{1em} 
	\begin{subfigure}{0.39\textwidth} 
		\includegraphics[width=\textwidth]{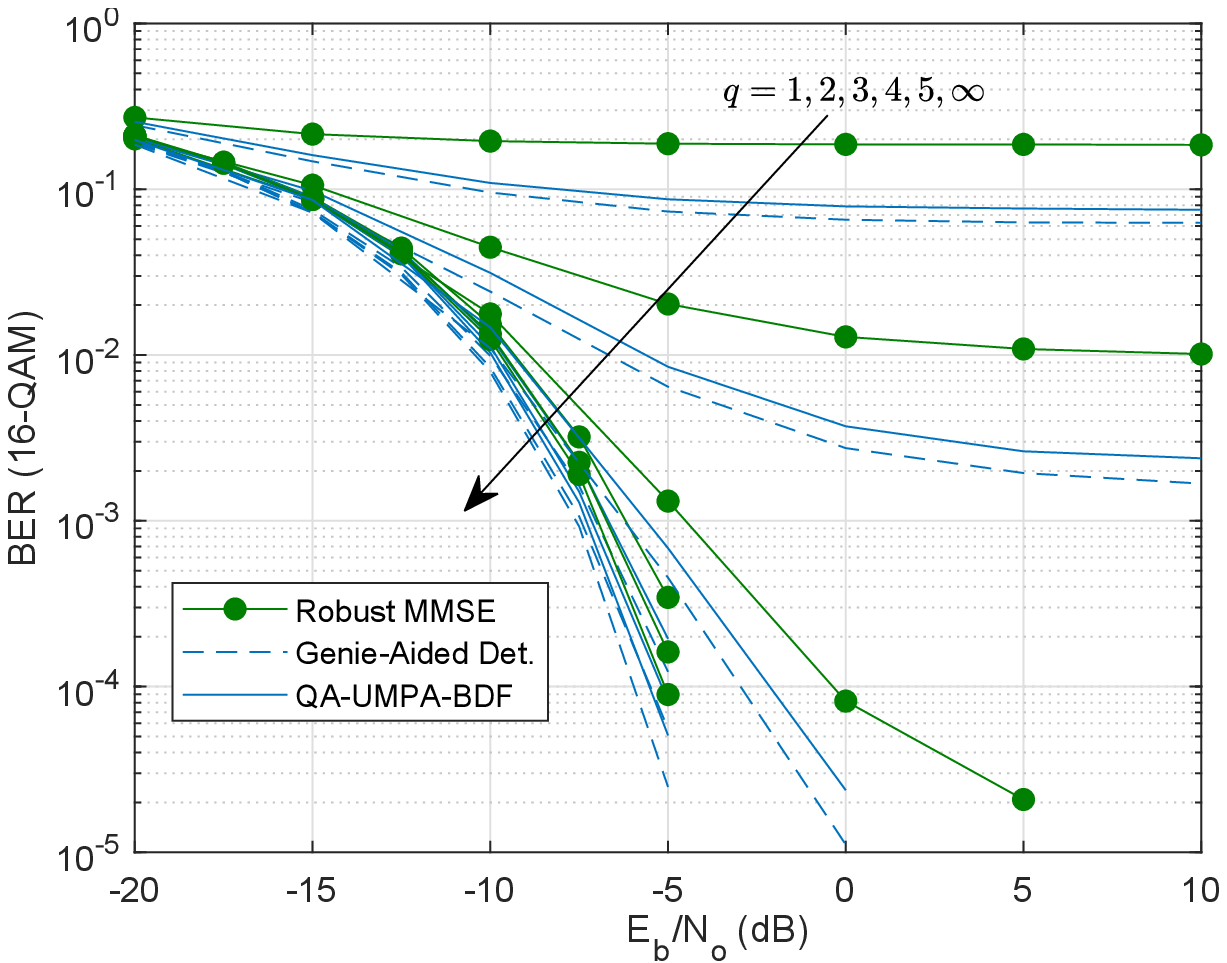}
		\caption{16-QAM.} 
		\label{fig:BER_SNR_16QAM_all_b}
	\end{subfigure}
	\caption{BER vs. SNR $K=15$, $q=1,2,3,4,5,\infty$ for QPSK (a), 16-QAM (b).}
	\label{fig:BER_SNR_QPSK_16QAM_all_b} 
\end{figure}

In the next simulation setting, the BER performances are observed for various modulation sizes (16-QAM, 8-PSK, 4-PSK and BPSK) in Fig.\ref{fig:BER_SNR_var_mod_q1_2}. Again, QA-UMPA-BDF has better performance compared to the benchmark detector for all modulation types, with significant performance difference for 16-QAM modulation. The reason to observe different BER for BPSK and QPSK is due to the correlation in the noise statistics stemming from the nonlinear quantizer. Such BER performance difference between BPSK and QPSK under quantization is also reported in \cite{ber_bpsk_qpsk_quant}.
\begin{figure}[htbp]
	\centering
	\begin{subfigure}{0.39\textwidth} 
		\includegraphics[width=\textwidth]{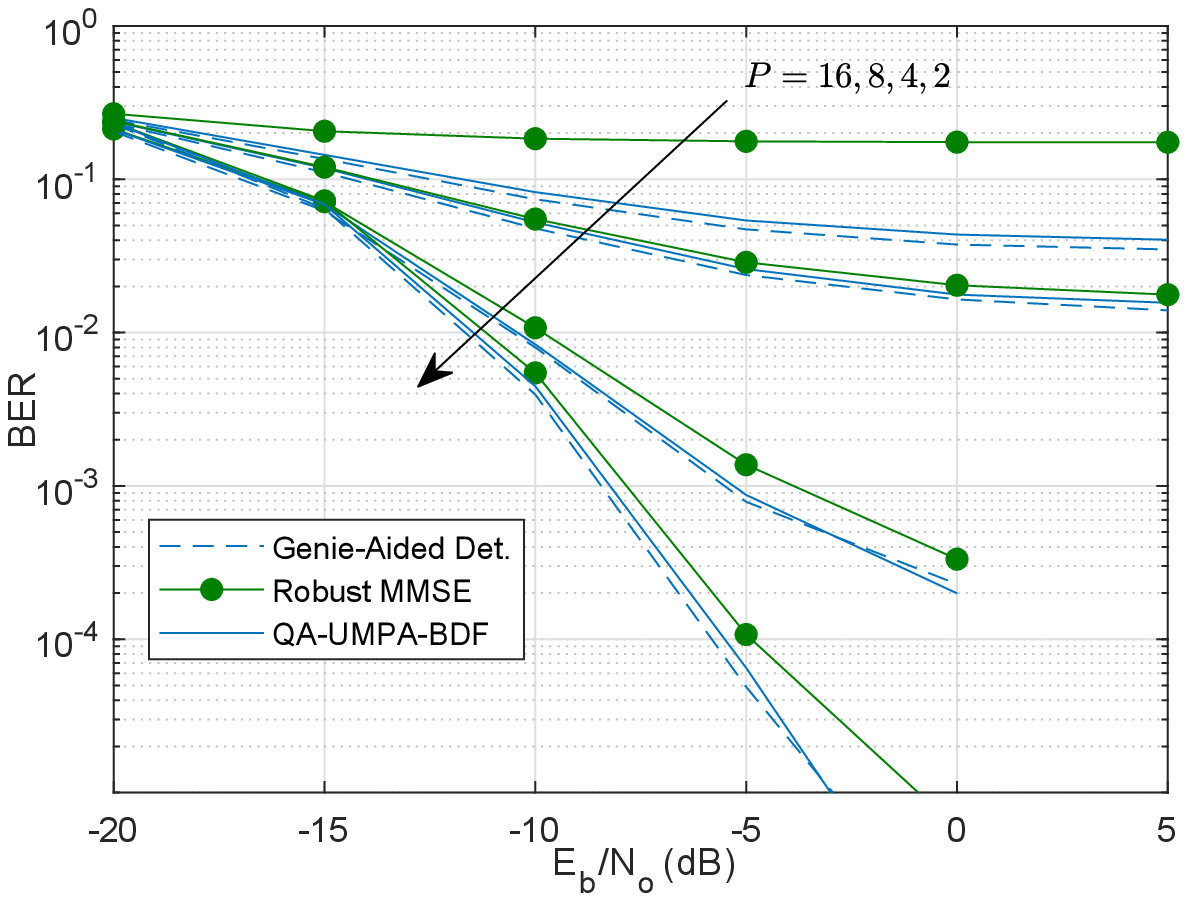}
		\caption{$q=1$.} 
		\label{fig:BER_SNR_var_mod_q1}
	\end{subfigure}
	\hspace{1em} 
	\begin{subfigure}{0.39\textwidth} 
		\includegraphics[width=\textwidth]{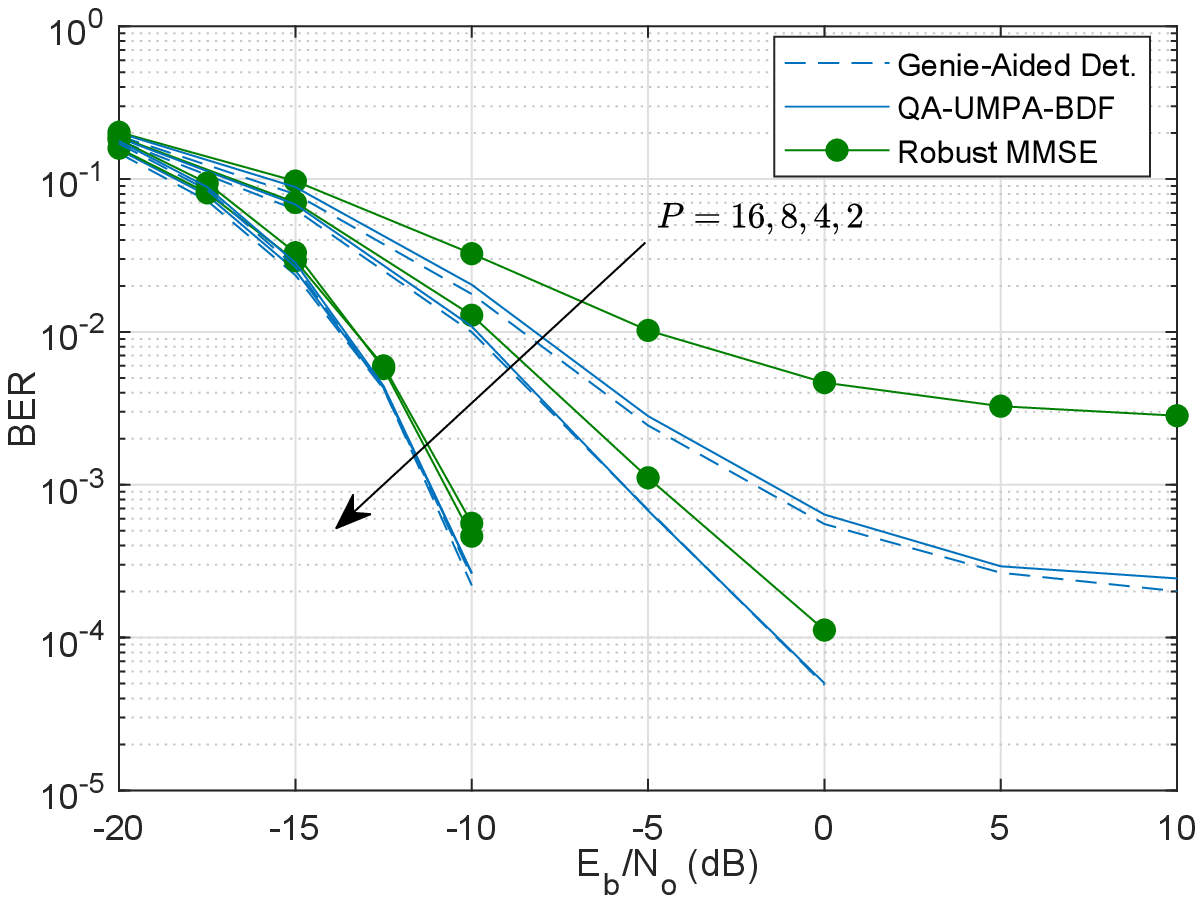}
		\caption{$q=2$.} 
		\label{fig:BER_SNR_var_mod_q2}
	\end{subfigure}
	\caption{BER vs. SNR for $P=16,8,4,2$, $K=10$, $q=1$ (a) and $q=2$ (b).}
	\label{fig:BER_SNR_var_mod_q1_2} 
\end{figure}

We also obtain per user AIR vs. SNR curves for $b=1$ and $b=2$ in Fig.~\ref{fig:AIR_SNR_b_1_2}. In Fig.~\ref{fig:AIR_SNR_b_1}, QA-UMPA-BDF detector asymptotically provides an AIR about $2.8$ bit per channel use (bpcu) for $q=1$ with 8-PSK, close to the maximum AIR of $3$ bpcu for 8-PSK. With 64-QAM, AIR can be asymptotically up to 3.5 bpcu for $q=1$. For $q=2$, we can see from Fig.~\ref{fig:AIR_SNR_b_2} that up to $5.5$ bpcu can be achieved with 64-QAM, close to the maximum AIR value of $6$ bpcu for 64-QAM, implying that a proper code with rate $5.5/6$ can provide very small BER values for 64-QAM. 

\begin{figure}[htbp]
	\centering
	\begin{subfigure}{0.40\textwidth} 
		\includegraphics[width=\textwidth]{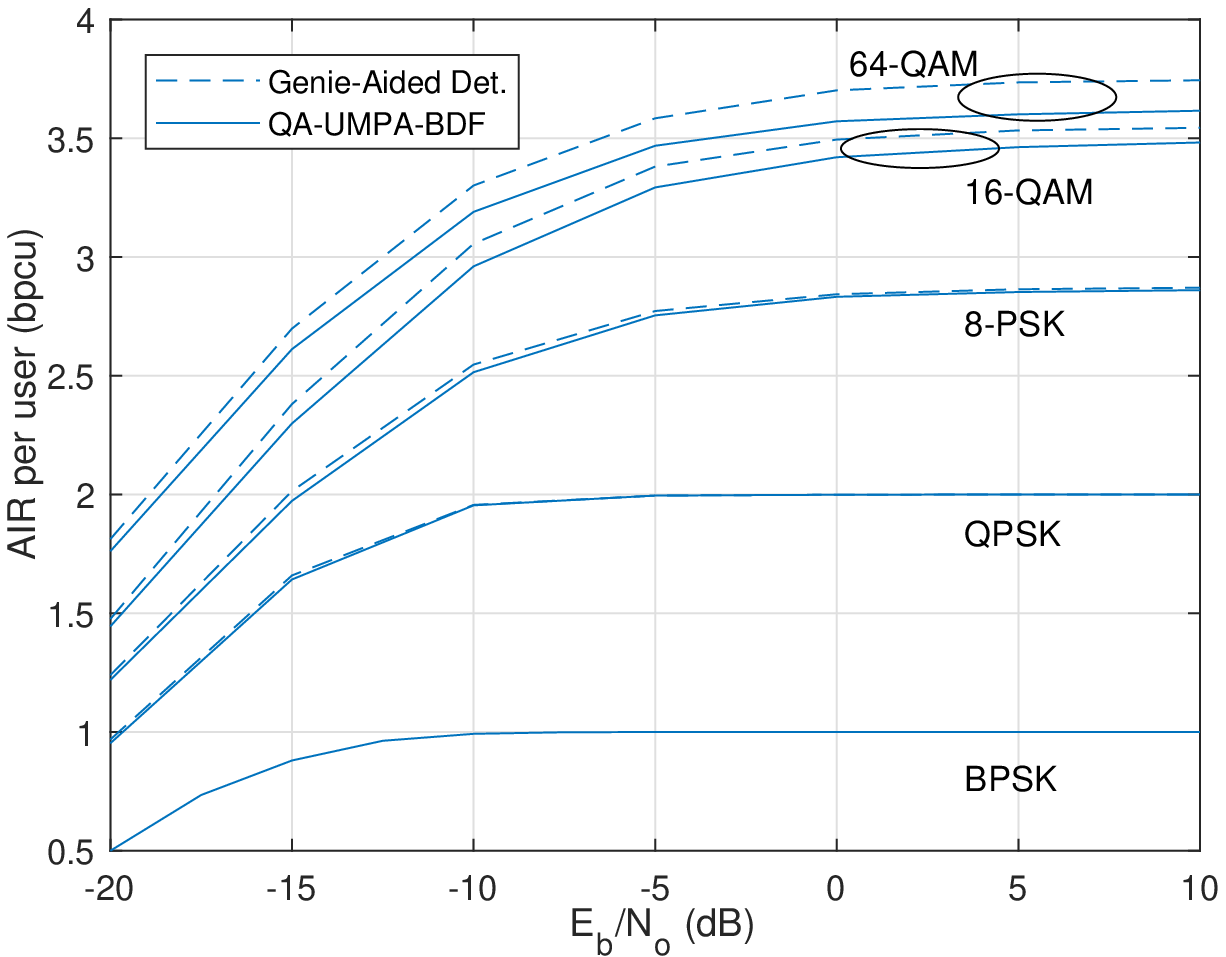}
		\caption{$b=1$.} 
		\label{fig:AIR_SNR_b_1}
	\end{subfigure}
	\hspace{1em} 
	\begin{subfigure}{0.41\textwidth} 
		\includegraphics[width=\textwidth]{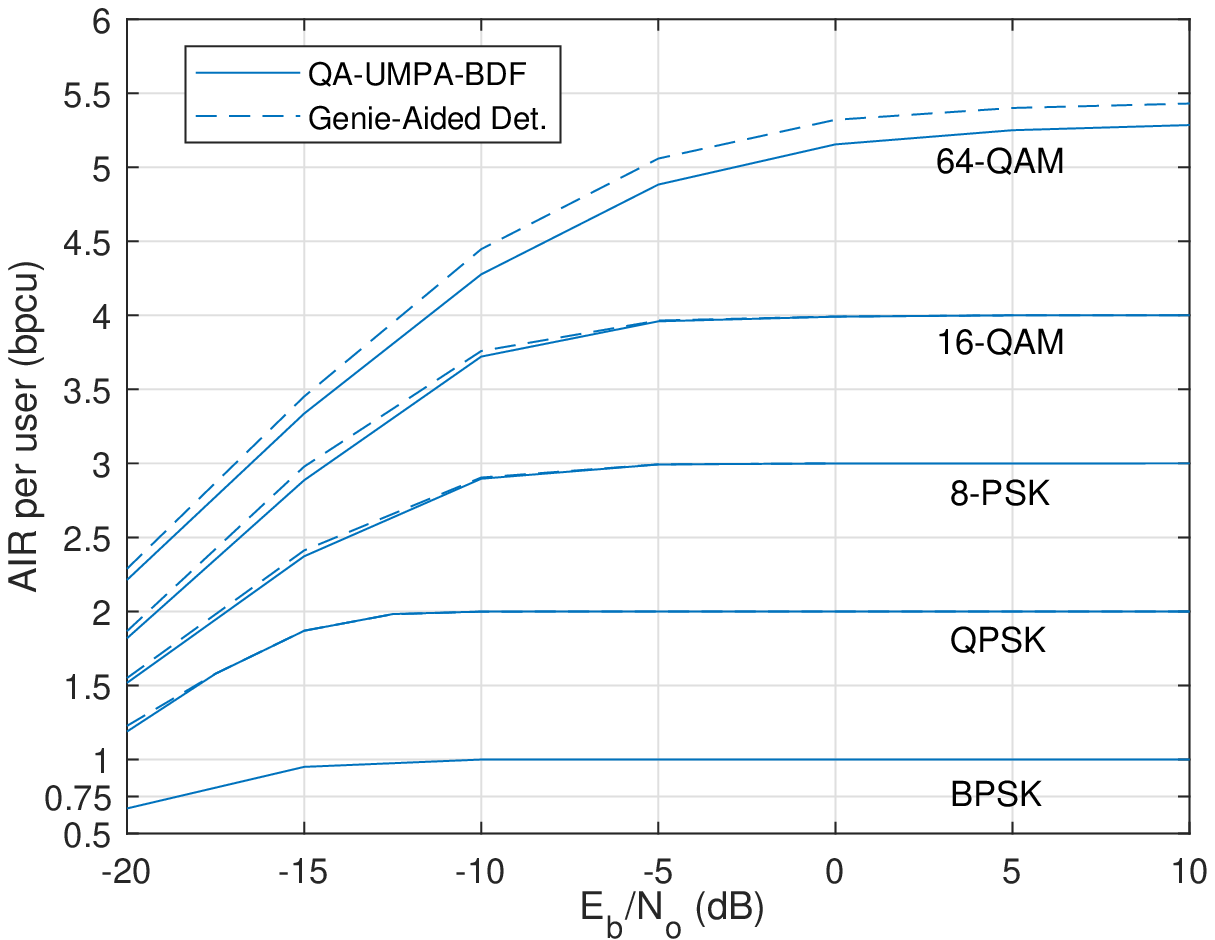}
		\caption{$b=2$.} 
		\label{fig:AIR_SNR_b_2}
	\end{subfigure}
	\caption{Per user AIR vs. $E_b/N_o$, $K=10$, $b=1$ (a), $b=2$ (b).}
	\label{fig:AIR_SNR_b_1_2} 
\end{figure}

In the next simulation setting, we obtain the total AIR instead of per user AIR vs. the number of users ($K$) in Fig.~\ref{fig:Total_AIR_all} when $E_b/N_o=0$ dB, $M=50$, $I=7$, $L=128$ with uniform power-delay profile as a challenging ISI channel scenario.
\begin{figure}[htbp]
\centering
\includegraphics[width=0.73\columnwidth]{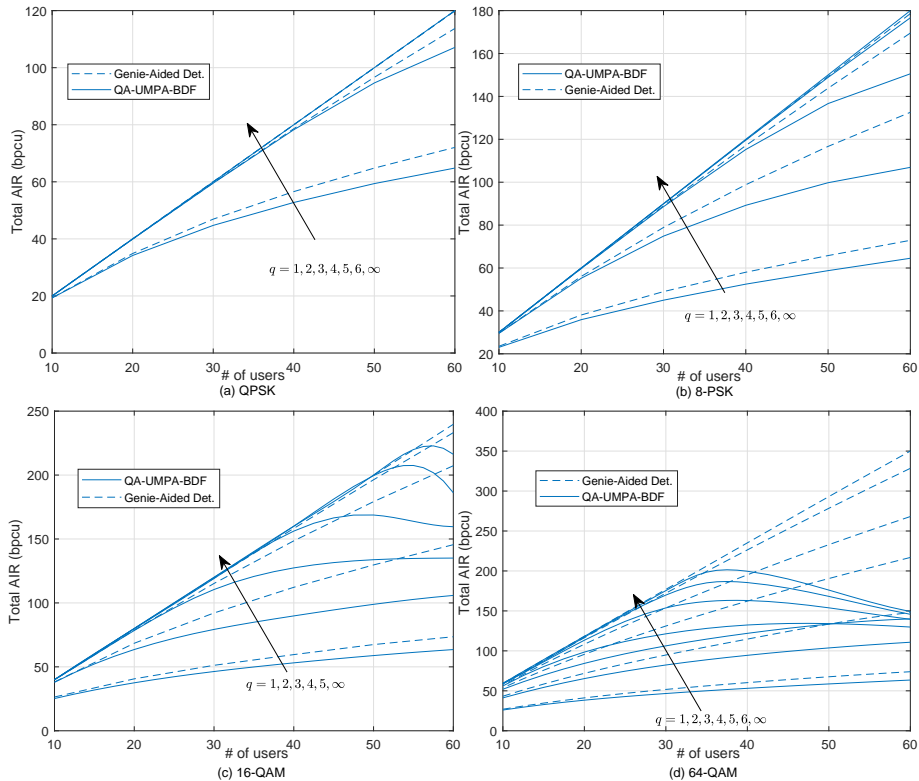}
\caption{Total AIR vs. number of users, $M=50$, $E_b/N_o=0$ dB, $q=1,2,\ldots,6,\infty$, QPSK (a), 8-PSK (b), 16-QAM (c), 64-QAM (d), $L=128$, uniform power-delay profile.} 
\label{fig:Total_AIR_all}
\end{figure}
The key takeaways from Fig.~\ref{fig:Total_AIR_all} are as follows:
\begin{itemize}
\item Strong total AIR performance is observed with QPSK even with $q=2$ and a very loaded case of 60 users, which is more than the number of antennas (maximum possible total AIR for 60 users is 120 bpcu with QPSK). Total AIR always rises with increasing $K$.
\item For 8-PSK, total AIR is better than QPSK for all $q$, if not similar. For $q=4$ maximum total AIR is achieved even with $K=60$. Total AIR always rises with increasing $K$.
\item For $q<4$ with 16-QAM, total AIR always increase with $K$. For $q=4, 5, \infty$, the maximum total AIR is observed for $K\approx 45, 53, 57$. Competitive performance with no degradation for up to about $55$ users is observed for $q>4$. Depending on the number of bits and users, 16-QAM has better total AIR performance than QPSK or 8-PSK in many cases.
\item 64-QAM has better total AIR performance compared to other modulation types for $q>4$ and $K<35$. For higher $K$ and lower $q$, smaller modulation orders provide better total AIR in some cases. The maximum total AIR of about $200$ bpcu is similar to that of 16-QAM.
\end{itemize}

For the final simulation cases, we present the BER performance for $E_b/N_o=0$ when the number of channel taps are varied in Fig.~\ref{fig:var_chan_tap} and the AIR per user performance vs. number of ADC bits for $E_b/N_o=0$ dB and $I=7$ in Fig.~\ref{fig:AIR_vs_no_bits}. 
\begin{figure}[htbp]
	\centering
	\begin{subfigure}{0.44\textwidth} 
		\includegraphics[width=\textwidth]{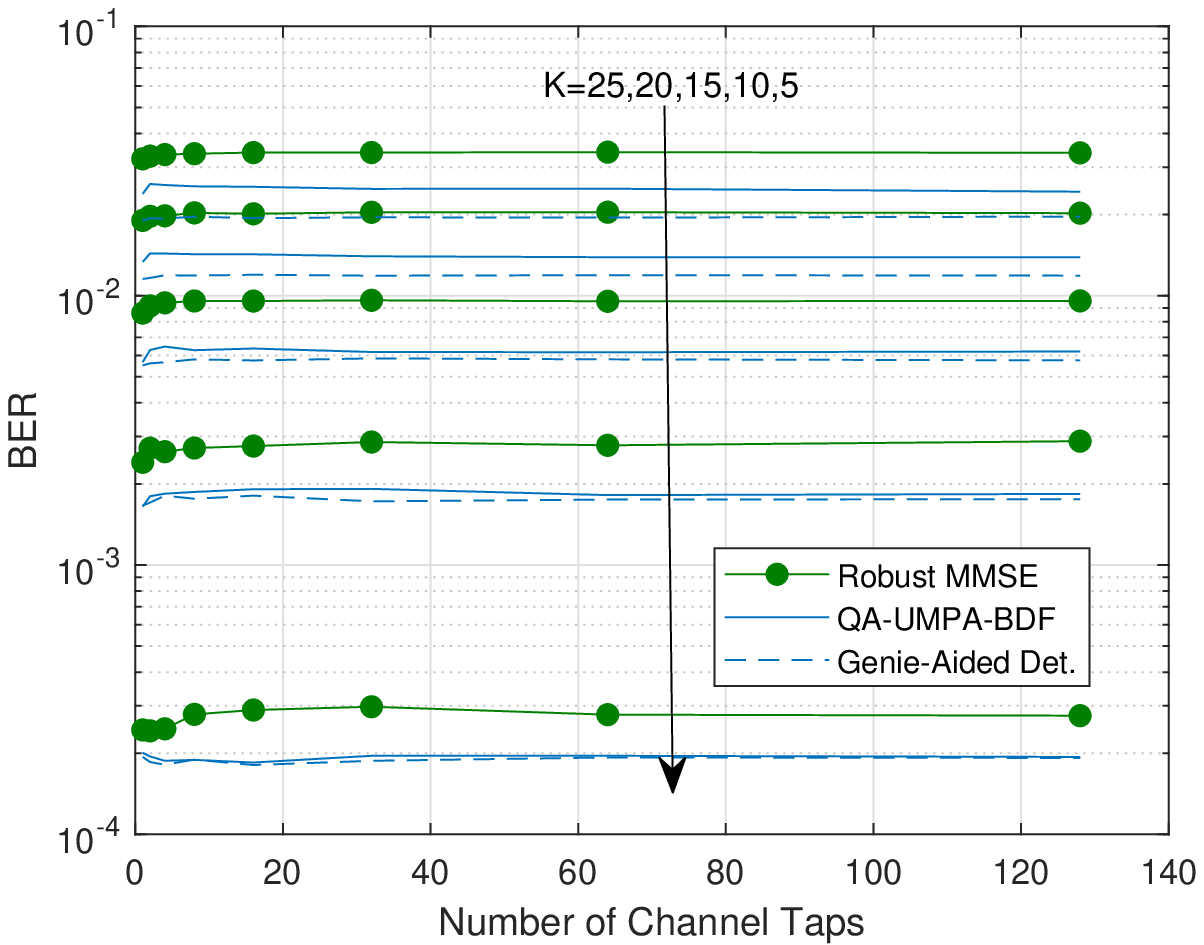}
		\caption{BER vs. $L$ for $E_b/N_o=0$ dB, $b=1$, QPSK, uniform power-delay profile.} 
		\label{fig:var_chan_tap}
	\end{subfigure}
	\hspace{1em} 
	\begin{subfigure}{0.44\textwidth} 
		\includegraphics[width=\textwidth]{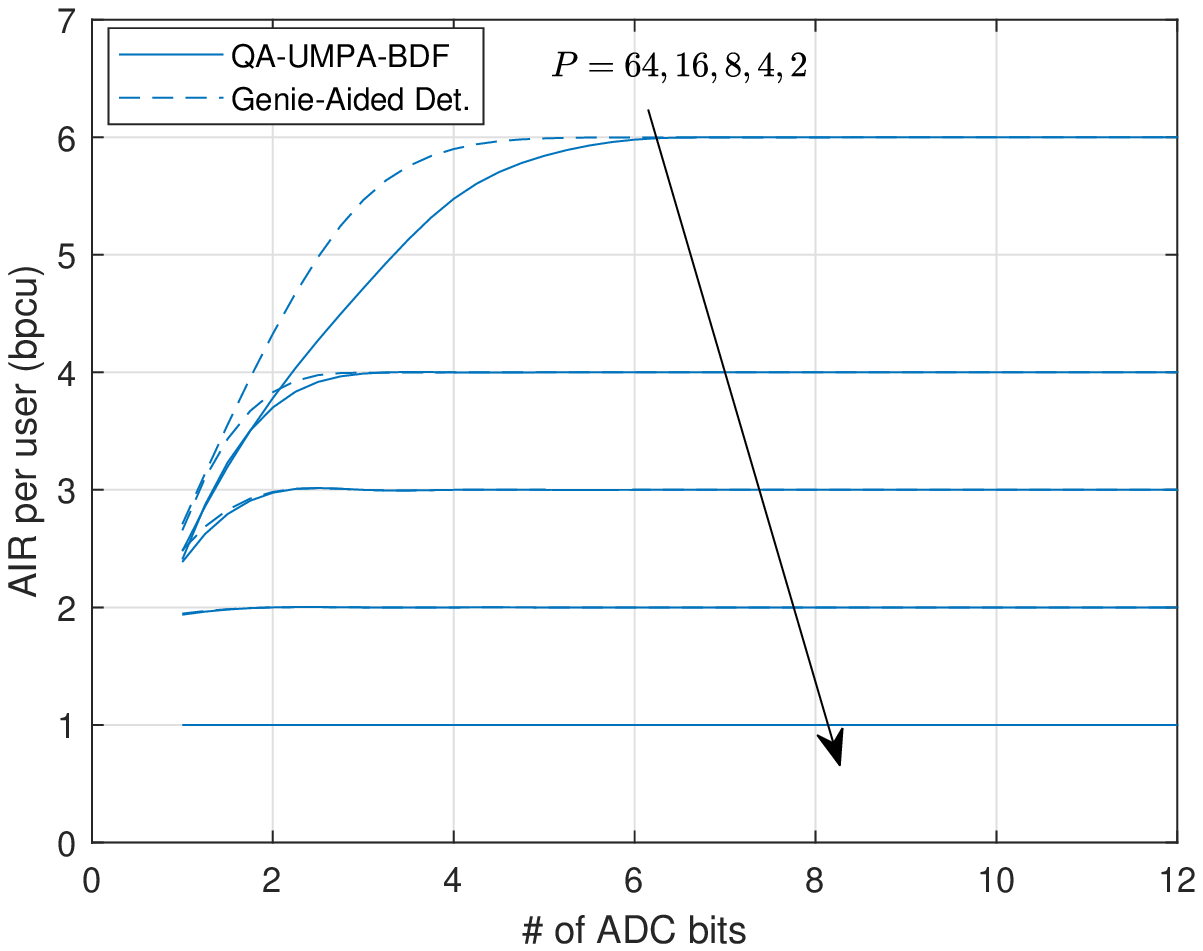}
		\caption{Per user AIR vs. number of bits $q$ for $K=25$, $E_b/N_o=0$ dB.} 
		\label{fig:AIR_vs_no_bits}
	\end{subfigure}
	\caption{BER vs. number of channel taps $L$ (a) or per user AIR vs. number of bits $q$ (b).}
	\label{fig:last_fig} 
\end{figure}
In Fig.~\ref{fig:var_chan_tap}, it can be seen that the proposed QA-UMPA-BDF detector has a very robust BER performance to the changes in the number of channel taps. It can cancel ISI in time-domain effectively even when the number of channel taps is as large as $128$, with no significant additional complexity ($J=0$). It also always has better performance compared to the representative detector. From Fig.~\ref{fig:AIR_vs_no_bits}, it can be stated that 64-QAM can be used with maximum possible AIR if $q>5$, while the maximum possible AIR is achieved for $q>2$ with 16-QAM. QA-UMPA-BDF provides higher AIR per user values for 64-QAM if $q>2$ for $K=25$.  If $q\leq2$, 64-QAM can be used as with outer channel coding as its AIR per user is better than the other modulation types.
    
\section{Conclusions}
In this paper, we proposed an LMMSE channel estimation and a low-complexity quantization-aware message passing detector based on bidirectional decision feedback. The proposed detector has very low complexity compared to the existing work in the literature for highly dispersive channels with large number of channel taps, thanks to its reduced state sequence estimation capability. Under imperfect CSI, the proposed QA-UMPA-BDF detector is observed to outperform (significantly for some cases) a representative detector from the literature with comparable complexity but lower spectral efficiency due to its requirement to use CP, for all examined cases.

\bibliographystyle{myIEEEtran}
\bibliography{IEEEabrv,referans}

\end{document}